\documentclass[aps,pra,preprint,onecolumn,superscriptaddress]{revtex4-1}
\usepackage{amssymb,amsmath}
\usepackage{bm}
\usepackage{subfigure}
\usepackage{setspace}
\usepackage{booktabs}
\usepackage{mathrsfs}
\usepackage{graphicx}
\usepackage{color,soul}
\usepackage{xr}

\def\ben{\begin{equation}}
\def\een{\end{equation}}

\newcommand{\mr}[1]{\mathrm{#1}}
\newcommand{\mc}[1]{\mathcal{#1}}

\newcommand{\mb}[1]{\mathbf{#1}}
\newcommand{\ud}{\mr{d}}
\newcommand{\ui}{\mr{i}}

\begin{document}

\title{Excited-State Response Theory Within the Context of the Coupled-Cluster Formalism}

\author{Mart\'in A. Mosquera}
\email{martinmosquera@montana.edu}
\affiliation{Department of Chemistry and Biochemistry, Montana State University, Bozeman, MT 59718,
USA}

\begin{abstract}  
Time-dependent response theories are foundational to the development of algorithms that determine
quantum properties of electronic excited states of molecules and periodic systems. They are employed
in wave-function, density-functional, and semiempirical methods, and are applied in an incremental
order: linear, quadratic, cubic, etc. Linear response theory is known to produce electronic
transitions from ground to excited state, and vice versa. In this work, a linear-response approach,
within the context of the coupled cluster formalism, is developed to offer transition elements
between different excited states (including permanent elements), and related properties. Our
formalism, second linear response theory, is consistent with quadratic response theory, and can
serve as an alternative to develop and study excited-state theoretical methods, including pathways
for algorithmic acceleration. This work also formulates an extension of our theory for general
propagations under non-linear external perturbations, where the observables are given by linked
expressions which can predict their time-evolution under arbitrary initial states and 
could serve as a means of constructing general state propagators. A connection
with the physics of wavefunction theory is developed as well, in which dynamical cluster operator
amplitudes are related to wavefunction linear superposition coefficients.
\end{abstract}

\maketitle
\section{Introduction}
Predicting the dynamics of electronic quantum systems, and ensembles of these, is a primary goal in
theoretical science for the understanding and discovery of cutting-edge physical and chemical
effects \cite{nelson2020non,park2020multireference,matsika2021electronic}. Without demanding
parameters besides the fundamental physical constants, quantum mechanics (QM) provides all the
necessary tools to determine all quantities needed for the theoretical modeling of quantum
phenomena.  This has led to the development of theoretical methods and algorithms that compute
observables connected to excited states, including development of quantum
\cite{anand2022quantum,ryabinkin2018qubit,romero2018strategies,xia2020qubit,tilly2020computation}
and machine-learning
\cite{smith2019approaching,keith2021combining,matsika2021electronic,dral2021molecular} technologies.
Such algorithms are often based on wavefunction or density functional theory, but they could also
rely on semiempirical theory, depending on their foundation their range of application varies.
There is a growing interest by the scientific community in excited-state phenomena linked to quantum
information science
\cite{higgott2019variational,bhattacharya2013thermodynamical,troiani2011molecular,saffman2010quantum},
quantum light emission/absorption
\cite{eshun2022entangled,fujihashi2021achieving,chen2021vibrational,parzuchowski2021setting,varnavski2020two,ma2021nonlinear,varnavski2022quantum},
cavity quantum dynamics \cite{mirza2022dissipative}, and multiphoton processes \cite{ou2007multi}.
Hence, quantum methods to compute properties connected to the modeling and understanding of these
phenomena can benefit from advanced theoretical tools.

Because of their balanced accessibility and computational power requirements, algorithms based on
linear response (LR) time-dependent density functional theory (TDDFT) are commonly used to
study the behavior of electrons subject to external perturbations (such as a low-intensity laser
field). LR TDDFT techniques
\cite{maitra2016perspective,laurent2013td,casida2012progress,maitra2021double}, through a single
matrix diagonalization, provide excited-state energies, and ground-to-excited-state multipolar
transition elements \cite{C95}, but other similar quantities can be computed as well. These methods
are quite suited for excited states mainly composed of single-electron transitions
\cite{elliott2011perspectives}. Excited states that originate from the simultaneous excitation of
two or more electrons are challenging to determine numerically. This also includes the study of multireference
states \cite{Fromager2007,Sharkas2012,wilbraham2017multiconfiguration}. Multireference theory 
\cite{roos1980complete,olsen2011casscf,olsen1988determinant,siegbahn1981complete}, due to its
widespread applicability to systems of strongly-correlated character, is to-date very actively motivating
the development of expanded methods that could stimulate newer generations of algorithms
\cite{ramakrishnan2015electronic}, which may also encompass density functional techniques. 

On the other hand, response theories within the context of wavefunction theory deliver information
as the aforementioned techniques
\cite{koch1990coupled,monkhorst1977calculation,dalgaard1983some,koch1991analytical,pedersen1997coupled,pedersen1997coupled,nascimento2019general}.
These demand higher computational power over DFT-based methods, but they are essential due to their
natural reliability and improvability. Wave-function/Green's function response theories have also
been extended to the multireference case
\cite{chattopadhyay2000development,samanta2014excited,jagau2012linear}.  Excited-state methods,
derived from response theory, that directly diagonalize a Hamiltonian are of general broad use as
they can be computationally convenient. An example of this is the well-known Bethe-Salpeter
equation.  \cite{vorwerk2019bethe,blase2020bethe,blase2020bethe}, capable of yielding
highly-accurate absorption spectra of extended systems and explain spectroscopic features seen in a
vast family of experiments. Similarly, multireference coupled cluster (MRCC) theory is among the
most advanced tools being developed currently to obtain high accuracy in energetics and
wave-function derived properties
\cite{evangelista2018perspective,jeziorski2010multireference,maitra2012unitary,hanrath2008multi,hanauer2011pilot,kohn2013state}.
MRCC methods are remarkably promising because they intengrate both dynamic- and strong-correlation
effects. So if activated for large systems, they would be quite beneficial.

This work presents the formulation of an extended linear-response approach, within single-reference
standard (non-Hermitian) coupled cluster theory
\cite{coester1958bound,coester1960short,vcivzek1966correlation,vcivzek1969use,bartlett2007coupled,emrich1981extension,
emrich1981extension2,monkhorst1977calculation,mukherjee1979response,ghosh1984use,stanton1993equation,zhang2019coupled},
that leads to the calculation of excited-state properties. This theory relies on a modification to
the initial state wavefunction of the system so one can extract properties of excited states through
linked coupled cluster (CC) equations. These are quantities such as matrix elements to study
transitions between excited states, as well as permanent dipoles of such states. This formulation is
based on an alternative linear response theory we developed previously, dubbed second linear
response theory (SLR)
\cite{mosquera2016sequential,mosquera2021second,mosquera2021second,kang2020efficient}. We have
applied it before within the context of time-dependent (TD) density functional theory to organic
semiconductors. The general working principle is founded on exact QM identities, and is applicable
to wave function methodologies, as shown in this work, where we develop an SLR approach within the
CC formalism and show that it provides excited-state expressions that are fully consistent with
established quadratic response theory. Then, we show SLR theory can be used to compute wave function
amplitudes in the linear regime where the electronic system is initially described by an
excited-state wavefunction. Finally, we extend our SLR theory to the non-linear case, where
excited-state information can be extracted from the analysis of generalized time-dependent
transition elements. This generalization, which is exact in principle, includes the description of
the evolution of an observable starting from an arbitrary initial state, such as a linear
superposition of different quantum states. The formalisms we present in this work could be used to
further expand the capabilities of response theories in theoretical and numerical contexts, where a
different angle on the fundamental problem of wave-function propagation can stimulate further
developments in the pursue of accuracy or to accelerate wavefunction-based algorithms to compute
excited-state properties.

\section{Definitions and Connection to Standard Linear Response Theory}
For any operator $\hat{\Omega}$
we write $\bar{\Omega}=\exp(-\hat{T})\hat{\Omega}\exp(+\hat{T})$, where $\hat{T}$ refers to the
standard ground-state cluster operator, which is assumed to be given. The symbol $\hat{\Omega}_{\mr{N}}$
denotes the normal-ordered form of $\hat{\Omega}$, {\slshape i.e.},
$\hat{\Omega}_{\mr{N}}=\{\hat{\Omega}\}$; also, we use the notation $\bar{\Omega}_{\mr{N}}=\{\bar{\Omega}\}$.
The letter $\mu$ labels transitions from the (single) ground-state reference of any order: singles,
doubles, triples, etc. So $\hat{\tau}_{\mu}$ is a product of electron-hole creation operators, and
$\hat{\tau}_{\mu}^{\dagger}$ its Hermitian conjugate. We use: i), $|0\rangle$ to refer to the reference
Hartree-Fock wavefunction, ii), $\langle \hat{\Omega}\rangle_0=\langle 0|\hat{\Omega}|0\rangle$,
and, iii), $\partial_t$ as a compact symbol for the partial derivative operator $\partial/\partial_t$.

The (non-relativistic) TD Hamiltonian of interest in this work is:
\ben
\hat{H}(t)=\hat{H}_{0}-f(t)\hat{B}
\een
where $\hat{H}_0$ is the static component, consisting of the kinetic, external (electron-nuclei interaction), and
electron-electron repulsion energies. The term $f(t)$ denotes the scalar driving potential the system is
subject to, and $\hat{B}$ the observable operator that couples to that potential. 
In addition, we are also interested in the evolution of an additional operator, denoted $\hat{A}$. Hence,
\ben
\langle A(t)\rangle = \langle [\hat{L}_0+\hat{\lambda}(t)]
e^{-\hat{x}(t)}\bar{A}e^{+\hat{x}(t)}\rangle_0
\een
where the operator $\hat{L}_0$ gives the left expression for the ground-state, $\langle
0|\hat{L}_0$. In terms of the well-known {\slshape lambda} operator this gives $\hat{L}_0=1+\hat{\Lambda}$.
The excitation operators read: $\hat{x}(t) = \sum_{\mu} x_{\mu}(t) \hat{\tau}_{\mu}$, and
$\hat{\lambda}(t)=\sum_{\mu} \hat{\tau}_{\mu}^{\dagger} \lambda_{\mu}(t)$, with $x_{\mu}(t)$ and
$\lambda_{\mu}(t)$ being the excitation/de-excitation TD amplitudes.
For the application of SLR theory, the above expression remains the starting point. But the initial
conditions of the $\hat{\lambda}(t)$ and $\hat{x}(t)$ terms are different, as we detail in
Section \ref{slr_theory}.

In this TD CC response formalism the left ket of the TD wavefunction is represented as:
\ben
\langle \Upsilon(t)|=\langle
0|[\hat{L}_0+\hat{\lambda}(t)]\exp[-\hat{x}(t)-\hat{T}+\ui \phi(t)]
\een 
where $E_0$ is the ground state energy, $\langle \hat{L}_0\bar{H}_0\rangle_0$, and $\phi(t)$ a TD
phase. The right ket reads
\ben
|\Phi(t)\rangle = \exp[\hat{T}+\hat{x}(t)-\ui \phi(t)]|0\rangle
\een
Using normal-ordering, we can express the Hamiltonian as:
$
\bar{H}(t) = E_0+\bar{H}_{0,\mr{N}}+\bar{v}(t)
$
where $\bar{v}(t)=-f(t)\bar{B}$.

In an ideal CC calculation both the left and right kets solve the full TD Schr\"odinger equation. In
practice, however, the differences between $|\Psi_{\mr{L}}(t)\rangle$ and $|\Psi_{\mr{R}}(t)\rangle$
are responsible for the non-Hermitian nature of CC response theory. But they offer the
quite desirable property of size-extensiveness, required to study large molecular systems and
periodic structures.

The motion equations of the $\hat{\lambda}$ and $\hat{x}$ operators can be derived from
stationarizing the action functional:
\ben
\begin{split}
\mc{F}[\bm{\lambda}, \bm{x},\phi]&=
\int \ud t~\Big\langle
[\hat{L}_0+\hat{\lambda}(t)]\Big\{e^{\displaystyle-\hat{x}(t)}\bar{H}(t) e^{\displaystyle+\hat{x}(t)}
-\ui \vec{\partial}_t[\hat{x}(t)-\ui \phi(t)] \Big\}\Big\rangle_0\\
&=\int \ud t~\Big[\langle \Upsilon(t)|\hat{H}(t)|\Phi(t)\rangle-\ui\langle
\Upsilon(t)|\vec{\partial}_t|\Phi(t)\rangle\Big]
\end{split}
\een
The symbols $\bm{\lambda}$ and $\mb{x}$ refer to the ``history'' of the amplitudes $\{\lambda_{\mu}(t)\}$ and
$\{x_{\mu}(t)\}$, respectively, whereas $\vec{\partial}_t$ indicates the time derivative is applied
to the ket $|\Phi(t)\rangle$.
Variations with respect to $\lambda_{\mu}$ and $x_{\mu}$ give the well-established TD equations:
\ben\label{x_eq}
\ui\partial_t x_{\mu}(t) = \langle
\hat{\tau}^{\dagger}_{\mu}e^{-\hat{x}(t)}[\bar{H}_{0}+\bar{v}(t)]e^{+\hat{x}(t)} \rangle_0
\een
and
\ben\label{l_eq}
-\ui \partial_t\lambda_{\mu}(t) = \langle
[\hat{L}_0+\hat{\lambda}(t)]e^{-\hat{x}(t)}[\bar{H}_{0}+\bar{v}(t),\hat{\tau}_{\mu}]e^{+\hat{x}(t)}\rangle_0
\een
Using the solution to the two last equations and by demanding that $\mc{F}=0$, the phase function takes the form:
\ben\label{phi_eq}
\partial_t\phi(t) = \langle[\hat{L}_0+\hat{\lambda}(t)]
e^{-\hat{x}(t)}\bar{H}(t)e^{+\hat{x}(t)} \rangle_0
\een
Because it originates from an action functional, the phase factor we use 
is different from that employed in other TD CC response formalisms.
For convenience we define 
\ben
\Delta \phi(t)=\int_0^t\ud s~\langle [\hat{L}_0+\hat{\lambda}(s)]e^{-\hat{x}(s)}
[\bar{H}_{0,N}+\bar{v}(s)]e^{+\hat{x}(s)}\rangle_0
\een 
so $\phi(t)=E_0 t+\Delta \phi(t)$.
Even though this phase does not influence the calculation of observables, it is important for the
interpretation of the right/left wavefunctions.

Now we specialize the above equations to the standard form of linear
response theory, and then to the SLR case, Section \ref{slr_theory}. For the latter, however, we
consider few additional terms that are due to the different type of initial condition that we use. 
We start by linearizing the TD CC
equations with respect to $\hat{x}$, $\hat{\lambda}$, and $\hat{v}$. 
This gives the following equation for the excitation amplitudes:
\ben\label{eq_eleven}
\ui\partial_t x_{\mu}(t) = \Big\langle
\hat{\tau}^{\dagger}_{\mu} \Big(\bar{v}(t) + [\bar{H}_0,
\hat{x}(t)]\Big)\Big\rangle_0
\een
Now we define the following operators:
\ben
\bar{H}_{\tau,\mu}^0= [\bar{H}_{0},\hat{\tau}_{\mu}]
\een
and
\ben
\bar{v}_{\tau,\mu}(t)= [\bar{v}(t),\hat{\tau}_{\mu}]
\een
In general
$\bar{\Omega}_{\tau,\mu}=\bar{\Omega}\hat{\tau}_{\mu}-\hat{\tau}_{\mu}\bar{\Omega}$.

Using the above definitions we obtain the equation:
\ben\label{eq_lambda_t}
-\ui\partial_t\lambda_{\mu}(t)=\Big\langle
\hat{L}_0\Big(\bar{v}_{\tau,\mu}(t)+[\bar{H}_{\tau,\mu}^0,\hat{x}(t)]\Big)+\hat{\lambda}(t)\bar{H}^0_{\tau,\mu}\Big\rangle_0
\een
To derive the above result one uses the fact that $\langle \hat{L}_0
[\bar{H}_0,\hat{\tau}_{\mu}]\rangle_0 = 0$.
Let us introduce the matrix:
\ben
(\mc{A})_{\mu\nu} = \langle \hat{\tau}^{\dagger}_{\mu}\bar{H}^0_{\tau,\nu}\rangle_0
\een
Because this is a non-symmetric (square) matrix, there is a set of left and right
eigenvectors $\{\bm{\Lambda}^I,~\mb{X}^I\}$, and eigenvalues (excitation energies) $\{\Omega_I\}$ such that 
$\mc{A}\mb{X}^I=\Omega_I\mb{X}^I$ and
$\mc{A}^{\mr{T}}\bm{\Lambda}^I=\Omega_I\bm{\Lambda}^I$. 
Following the steps shown in the supporting material, we find the well-known linear response
expressions for the ground-to-excited state transition matrix elements:
\ben\label{eq_left}
\langle \Psi_I|\hat{A}|\Psi_0\rangle = \sum_{\mu}
\Lambda_{\mu}^I\langle\hat{\tau}^{\dagger}_{\mu}\bar{A}\rangle_0
\een
and
\ben\label{eq_right}
\langle\Psi_0|\hat{A}|\Psi_I\rangle = \sum_{\mu}\langle\hat{L}_0\bar{A}_{\tau,\mu}\rangle_0 X_{\mu}^I-
\sum_J \frac{F^{IJ}(\bm{\Lambda}^J\cdot \mb{\bar{A}})}{\Omega_I+\Omega_J}
\een
$F^{IJ}$ is the matrix element:
\ben
F^{IJ}=\sum_{\mu\nu} X_{\mu}^IF_{\mu\nu}X_{\nu}^J
\een
where $F_{\mu\nu} = \langle \hat{L}_0 [\bar{H}^0_{\tau,\mu},\hat{\tau}_{\nu}]\rangle_0$, 
and $\bm{\Lambda}^J\cdot \mb{\bar{A}} = \sum_{\mu} \Lambda^J_{\mu}\langle
\hat{\tau}^{\dagger}_{\mu}\bar{A}\rangle_0$.
This result holds for the observable $B$ as well. 

\section{Second Linear Response Theory}\label{slr_theory}

\begin{figure}[htb!]
\centering
\includegraphics[scale=0.35]{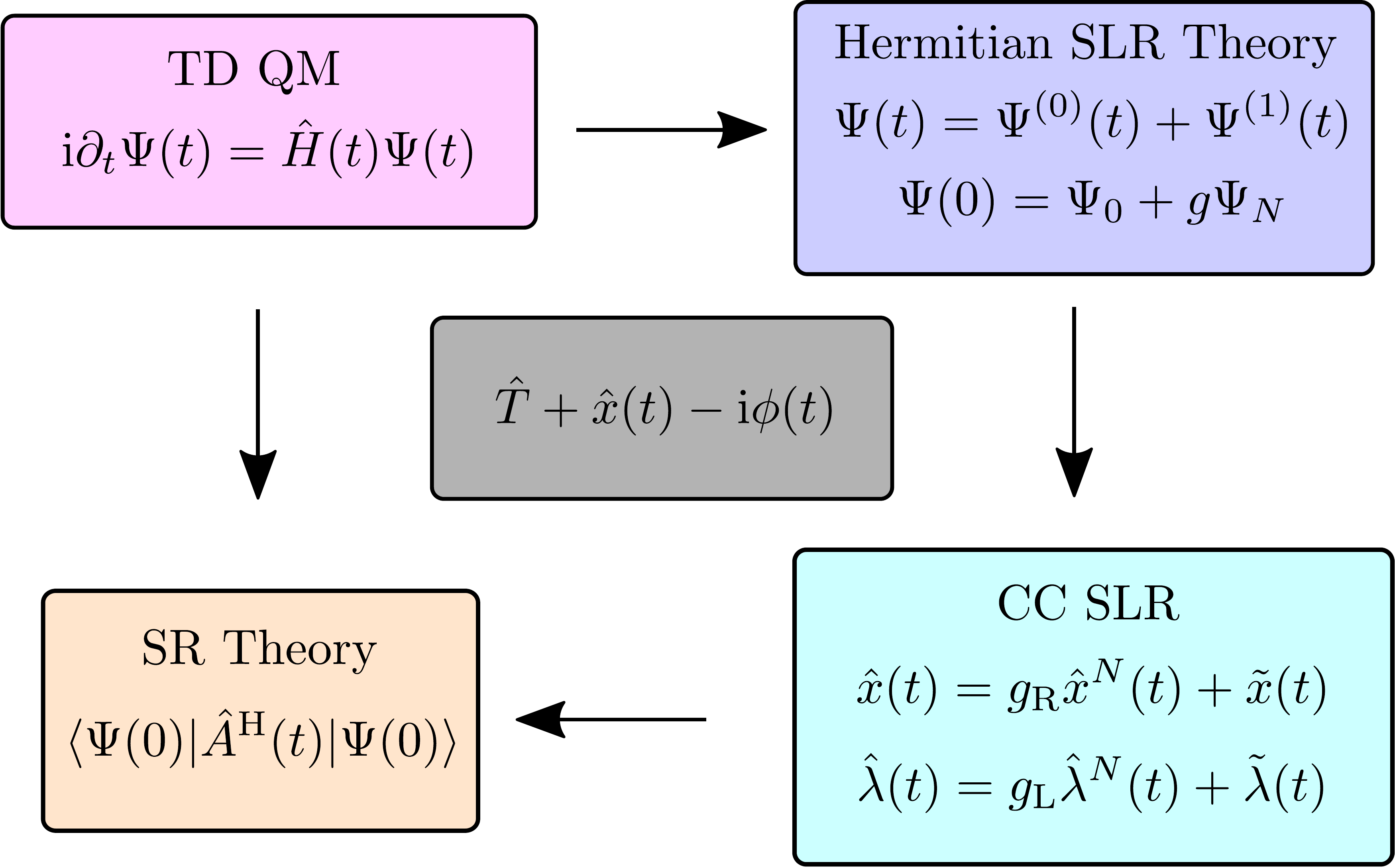}
\caption{Theoretical components explored in this work. Standard TD QM defines the
quantities that are to be represented by our CC approaches. The starting point is the use of LR
QM where the initial state is not the ground state, but a combination of its
ground-state wavefunction with an excited state of interest ($\Psi_N$). A LR CC theory is formulated
to cover this situation, and is then extended to consider cases beyond the linear response regime,
yielding SR theory, in which an observable is propagated for a general initial state (not purely ground
state). In the center frame we show the combined set of operators used to
examine the response of the system to external perturbations.}\label{thefig}
\end{figure}

In this section we develop an alternative formalism to compute excited-state transition elements.
We observe, as in the linear response case, that the left response vector contributes counter-
and clockwise elements, whereas the right vector does so only for counter-clockwise ones.  Although
we follow different theoretical steps, the matrix elements we predict are consistent with quadratic
response theory. We remark, however, that the phase expression we utilize differs from other
CC-based response theories. This phase does not affect the transition elements. 
However, as we show in Section \ref{waveamp}, our phase equation is useful
to interpret wavefunction amplitudes that emerge from our second response (SR) theory. The steps
followed are pictorically summarized in Fig. \ref{thefig}.

From standard quantum mechanics, we apply linear-response analysis to the case where the system is
initially described by a linear combination of the form: 
\ben
|\Psi(t=0;g)\rangle =
|\Psi_0\rangle+g|\Psi_N\rangle
\een
here $\Psi_N$ denotes an excited-state of interest. 
The linear-response TD WF is:
\ben\label{psi1}
|\Psi(t;g)\rangle=|\Psi^{(0)}(t;g)\rangle+|\Psi^{(1)}(t;g)\rangle
\een
where 
\ben\label{psi2}
|\Psi^{(0)}(t;g)\rangle=e^{-\ui\hat{H}_0 t}|\Psi(0;g)\rangle
\een
and
\ben\label{psi3}
|\Psi^{(1)}(t;g)\rangle = -\ui\int_0^t \ud s~e^{-\ui\hat{H}_0 (t-s)}\hat{V}(s)e^{-\ui\hat{H}_0
s}|\Psi^{(0)}(0;g)\rangle
\een
The response function now reads:
\ben
\mc{R}(\omega;g) = \int_{-\infty}^{+\infty}\ud t~e^{\ui(\omega\pm\ui\eta)t}\frac{\delta }{\delta
f(s)}\Big\{\langle \Psi^{(0)}(t;g)|\hat{A}|\Psi^{(1)}(t;g)\rangle+\mr{c.c.}\Big\}\Bigg|_{s=0,f=0}
\een
Using these equations and taking $\eta\rightarrow 0^+$, we find the following:
\ben
\begin{split}
\lim_{g\rightarrow 0} \frac{\partial}{\partial g} \mc{R}(\omega;g)
=-\sum_J&\Bigg[\frac{\langle\Psi_0|\hat{A}|\Psi_J\rangle\Big(\langle\Psi_J|\hat{B}|\Psi_N\rangle-
\delta_{JN}\langle\Psi_0|\hat{B}|\Psi_0\rangle\Big)}{\omega-\Omega_J}\\
&-\frac{\langle\Psi_J|\hat{A}|\Psi_0
\rangle\Big(\langle\Psi_N|\hat{B}|\Psi_J\rangle-
\delta_{NJ}\langle\Psi_0|\hat{B}|\Psi_0\rangle\Big)}{\omega+\Omega_J}\Bigg]
\end{split}
\een
Although we used a single variable ($g$) for the above equations, we now split the analysis into a
left and a right mathematical problem by using one superposition variable ($g_{\mr{L}}$) for the counter-clockwise
component and a second variable ($g_{\mr{R}}$) for the clockwise one, where variations with respect to either gives
the information of interest. Starting from Eqs. (\ref{psi1}-\ref{psi3}), we consider the wavefunctions $\langle \Psi(t;g_{\mr{L}})|$,
$|\Psi(t;g_{\mr{R}})\rangle$, and their zero and first order components.
For example:
$\langle \Psi^{(0)}(t=0;g_{\mr{L}})|=\langle \Psi_0|+g_{\mr{L}}\langle \Psi_N|$, where
$\langle \Psi^{(0)}(t;g_{\mr{L}})|=\langle \Psi^{(0)}(t=0;g_{\mr{L}})|\exp(\ui \hat{H}_0 t)$.
In a similar way we obtain the wavefunction $|\Psi^{(0)}(t;g_{\mr{R}})\rangle$. 

Henceforth, we introduce the function:
\ben
\begin{split}
R_2(\omega;g_{\mr{L}}, g_{\mr{R}}) = \int_{-\infty}^{+\infty}&\ud t~e^{\ui(\omega\pm\ui\eta)t}\frac{\delta }{\delta
f(s)}\Big\{\langle \Psi^{(0)}(t;g_{\mr{L}})|\hat{A}|\Psi^{(1)}(t;g_{\mr{R}})\rangle\\
&+\langle \Psi^{(1)}(t;g_{\mr{L}})|\hat{A}|\Psi^{(0)}(t;g_{\mr{R}})\rangle
\Big\}\Bigg|_{s=0,f=0}
\end{split}
\een
In agreement with the function $\mc{R}$, $R_2$ satisfies:
\ben
\lim_{\omega\rightarrow \Omega_I}\lim_{g_{\mr{L}},g_{\mr{R}}\rightarrow 0}
-(\omega-\Omega_I)\frac{\partial}{\partial g_{\mr{R}}}R_2 = 
\langle\Psi_0|\hat{A}|\Psi_I\rangle\Big(\langle\Psi_I|\hat{B}|\Psi_N\rangle-
\delta_{IN}\langle\Psi_0|\hat{B}|\Psi_0\rangle\Big)
\een
and
\ben
\lim_{\omega\rightarrow -\Omega_I}\lim_{g_{\mr{L}},g_{\mr{R}}\rightarrow 0}
(\omega+\Omega_I)\frac{\partial}{\partial g_{\mr{L}}}R_2 =
\langle\Psi_I|\hat{A}|\Psi_0\rangle\Big(\langle\Psi_N|\hat{B}|\Psi_I\rangle-
\delta_{NI}\langle\Psi_0|\hat{B}|\Psi_0\rangle\Big)
\een
We now proceed to solve the CC linear response equations under the initial condition where the system
is in a linear combination of the ground state and some excited state of interest. We label this
excited state as $N$. 

If the system is unperturbed then it must behave as a stationary state that satisfies the standard
linear response equations. Therefore we seek for a solution set as shown below:
\ben\label{slr_ansatz}
\begin{split}
\hat{x}(t;g_{\mr{R}}) &= g_{\mr{R}}\hat{x}^N(t)+\tilde{x}(t;g_{\mr{R}})\\
\hat{\lambda}(t;g_{\mr{L}},g_{\mr{R}}) &= g_{\mr{L}}\hat{\lambda}^N(t)+\tilde{\lambda}(t;g_{\mr{L}},g_{\mr{R}})\\
\phi(t;g_{\mr{L}},g_{\mr{R}})&=g_{\mr{R}}\phi^N(t)+\tilde{\phi}(t;g_{\mr{L}},g_{\mr{R}})
\end{split}
\een
The operators $\hat{x}^N(t)$, $\hat{\lambda}^N(t)$ and the phase $\phi^N(t)$ represent the
stationary state that would occur in the absence of an external perturbation ($\hat{v}(t)=0$).
The terms $\tilde{x}(t)$, $\tilde{\lambda}(t)$, and $\tilde{\phi}(t)$ are the ``new'' response
operators/phase, they provide information about the evolution of the system. We express the operators as
$\tilde{x}(t)=\sum_{\mu}\tilde{x}_{\mu}(t)\hat{\tau}_{\mu}$, and
$\tilde{\lambda}(t)=\sum_{\mu}\hat{\tau}^{\dagger}\tilde{\lambda}_{\mu}(t)$.
As we show later on, the operator $\tilde{\lambda}$ depends on both $g_{\mr{L}}$ and
$g_{\mr{R}}$, in addition to time.
For the phase we use the right amplitude $g_{\mr{R}}$ only as the operator $\hat{x}^N(t)$ determines this
object, besides $E_0$. Its response part, $\tilde{\phi}$, on the other hand, depends on $\tilde{x}$
and $\tilde{\lambda}$, and thereby on $g_{\mr{L}}$ and $g_{\mr{R}}$. 

The vectors $\hat{x}^N(t)$ and $\hat{\lambda}^N(t)$ stationarize their respective equations.
Equation (\ref{eq_eleven}) reads
\ben
\ui\partial_t x_{\mu}^N(t) = \sum_{\nu}\mc{A}_{\mu\nu}x^N_{\nu}(t)
\een
This indicates that $x_{\mu}^N(t) = X^N_{\mu}\exp(-\ui\Omega_N t)$. 
The vector $\lambda_{\mu}^N(t)$
follows a different relation:
\ben
-\ui g_{\mr{L}}\partial_t \lambda_{\mu}^N(t)=\Big\langle g_{\mr{R}}\hat{L}_0[\bar{H}^0_{\tau,\mu},\hat{x}^N(t)]
+g_{\mr{L}}\hat{\lambda}^N(t)\bar{H}^0_{\tau,\mu}\Big\rangle_0
\een
The solution to this equation when both $g_{\mr{L}}$ and $g_{\mr{R}}$ are different from zero is
not physically meaningful because $\hat{x}^N(t)$ introduces a counter-clockwise term, and by
extension contributions from all frequencies. Therefore, we are
interested in physical case where $g_{\mr{R}}=0$ and $g_{\mr{L}}\neq 0$, and then the limit
$g_{\mr{L}}\rightarrow 0$. Thus we take 
$\lambda^N_{\mu}(t)=\Lambda^N_{\mu}\exp(\ui\Omega_N t)$, which meets physical expectation.
The phase $\phi^N(t)$
satisfies:
\ben
\phi^N(t)=E_0 t+\Delta \phi^N(t)
\een
in the above equation $\Delta\phi^N(t)=\int_0^t \ud s~\langle\hat{L}_0 [\bar{H}_0,\hat{x}^N(s)]\rangle_0$.

To derive the linearized time-dependent equations that from Eqs.
(\ref{x_eq}-\ref{phi_eq}), we include terms that are proportional to $g_{\mr{L}}$ or $g_{\mr{R}}$ (for
example, a term like $g_{\mr{R}}[\bar{v}(t),\hat{x}^N(t)]$ needs to be included), as these two
numbers, from a linear response perspective, are fixed, and they remain non-zero after completing
the limiting procedures that we apply. Any term that is quadratic in $g_{\mr{L}}$ or $g_{\mr{R}}$ in
the weak perturbation limit is neglected because these vanish.

The SLR equation for the components of the operator $\tilde{x}$ reads:
\ben\label{slr_x}
\ui\partial_t \tilde{x}_{\mu}(t) = \Big\langle\hat{\tau}_{\mu}^{\dagger}\Big\{ 
  [\bar{H}_0,
  \tilde{x}(t)]+\bar{v}(t)+g_{\mr{R}}[\bar{v}(t),\hat{x}^N(t)]+g_{\mr{R}}\hat{M}(t)\Big\}\Big\rangle_0
\een
where
\ben
\hat{M}(t) = \Big[[\bar{H}_0,\hat{x}^N(t)],\tilde{x}(t)\Big]
\een
The conjugate operator $\tilde{\lambda}(t)$ follows the equation:
\ben\label{slr_l}
\begin{split}
-\ui\partial_t \tilde{\lambda}_{\mu}(t) = \Big\langle\hat{L}_0\big(\bar{v}_{\tau,\mu}(t)+
 &[\bar{H}^0_{\tau,\mu},\tilde{x}(t)]\big)+\tilde{\lambda}(t)\bar{H}^0_{\tau,\mu}
 +g_{\mr{R}}\hat{L}_0[\bar{v}_{\tau,\mu}(t),\hat{x}^N(t)]\\
 &+g_{\mr{L}}\hat{\lambda}^N(t)\bar{v}_{\tau,\mu}(t)+\hat{Q}_{\mu}(t)\Big\rangle_0
\end{split}
\een
where
\ben
\hat{Q}_{\mu}(t)=g_{\mr{R}}\hat{L}_0\Big[[\bar{H}^0_{\tau,\mu},\hat{x}^N(t)],\tilde{x}(t)\Big]+
g_{\mr{L}}\hat{\lambda}^N(t)[\bar{H}^0_{\tau,\mu},\tilde{x}(t)]+g_{\mr{R}}\tilde{\lambda}(t)[\bar{H}^0_{\tau,\mu},\hat{x}^N(t)]
\een
The SLR phase is given by:
\ben
\partial_t \tilde{\phi}(t) = (1-g_{\mr{R}}) E_0+\partial_t\Delta \tilde{\phi}(t)
\een
where
\ben
\begin{split}
\Delta\tilde{\phi}(t;g_{\mr{L}}, g_{\mr{R}})=&\int_0^t\ud s~
\Big\langle g_{\mr{L}}\hat{\lambda}^N(s)\bar{v}(s)+g_{\mr{L}}\hat{\lambda}^N(s)[\bar{H}_0,\tilde{x}(s)]+
g_{\mr{R}}\tilde{\lambda}(s)[\bar{H}_0,\hat{x}^N(s)]\\
&+\hat{L}_0\Big\{[\bar{H}_0,
  \tilde{x}(s)]+\bar{v}(s)+g_{\mr{R}}[\bar{v}(s),\hat{x}^N(s)]+g_{\mr{R}}\hat{M}(s)\Big\}\Big\rangle_0
\end{split}
\een
The last three SLR equations are fully consistent with standard LR when $g=0$.

For these SLR equations, it is important to note the initial conditions
$\tilde{\lambda}_{\mu}(t=0)=\tilde{x}_{\mu}(0)=0$, and this holds regardless of the values of
$g_{\mr{L}}$ and $g_{\mr{R}}$. After carrying out the mathematical analysis of the response
functions, as shown in the supporting material, we obtain the relation:
\ben\label{PsiIN}
\langle\Psi_I|\hat{B}|\Psi_N\rangle=\delta_{IN}\langle \hat{L}_0\bar{B}\rangle_0+\langle\hat{\Lambda}^I
\bar{B}_{X,N}\rangle_0
+\sum_J
\Bigg[\frac{C_{IN,J}}{\Omega_I-\Omega_J-\Omega_N}\Bigg]
(\bm{\Lambda}^J\cdot\bar{\mb{B}})
\een
where
\ben
C_{IN,J}=
\langle\hat{\Lambda}^{I}\big[[\bar{H}_0,\hat{X}^N],\hat{X}^J\big]\rangle_0
\een
and $\hat{X}^J=\sum_{\mu}X_{\mu}^J\hat{\tau}_{\mu}$, 
$\hat{\Lambda}^I = \sum_{\mu}\Lambda_{\mu}^I\hat{\tau}^{\dagger}_{\mu}$.
Both the left and right evaluations give the same element, one only has to swap the $N$ and $I$
indeces.

In the limit where the CC excited state problem is solved to all orders, the last term in Eq.
(\ref{PsiIN}) eliminates $\langle \hat{\Lambda}^I\hat{X}^N\hat{B}\rangle_0$, so the matrix element is
given by $\langle\hat{\Lambda}^I\bar{B}\hat{X}^N\rangle$. This implies that the last term in
Eq. (\ref{PsiIN}) is in such limit finite, but not necessarily otherwise. For this reason, it may be
important to apply a regularization scheme in case there is a term $\Omega_I-\Omega_J-\Omega_N$ that
is quite close to zero. Alternatively, as an additional approximation, not explored in this work, for the sake of eliminating
divergences one can neglect the difference $\Omega_I-\Omega_N$. It holds true for the case of
permanent-dipole determination, but not for transition elements. 

\section{Wavefunction Amplitudes}\label{waveamp}
Although the initial state we employed before is a quantum mixture of ground and excited state, one
can also analyze through such initial state the situation where the system begins evolving from the
excited state $N$, and the response to a weak perturbation can be determined. Note that
$
\partial/\partial g| \Psi^{(0)}(t=0;g)\rangle = |\Psi_N\rangle
$, where the first derivative of the initial of state with respect to $g$ gives the excited-state wavefunction.
When we apply the same operation to the first response wave function it is found that:
\ben
\frac{\partial}{\partial g}| \Psi^{(1)}(t;g)\rangle=
-\ui\int_0^t \ud s~e^{-\ui\hat{H}_0 (t-s)}\hat{V}(s)e^{-\ui\hat{H}_0
s}|\Psi_N\rangle
\een
This is equivalent to the result of applying standard linear response, where the initial state is entirely
described by $\Psi_N$.

Let us introduce the following expansion:
\ben
\partial_g|\Psi^{(1)}(t)\rangle=\sum_I \mc{C}_I(t)|\Psi_I\rangle
\een
where the amplitude $\mc{C}_I(t)$ is given by $ \mc{C}_I(t)=\langle
\Psi_I|\partial_g\Psi^{(1)}(t)\rangle $ ($\partial_g=\partial /\partial g$). This object then
describes contribution of state $I$ to the response of the initial excited state to a perturbation,
and it can be related to response CC coefficients. But before proceeding to show this, for a
function $h$ of the coefficients $g_{\mr{L}}$ and $g_{\mr{R}}$, the following notation is used:
\ben
\begin{split}
h_{\mr{r}}&=\lim_{g_{\mr{L}},g_{\mr{R}}\rightarrow 0} \frac{\partial h}{\partial g_{\mr{R}}}\\
h_{\mr{l}}&=\lim_{g_{\mr{L}},g_{\mr{R}}\rightarrow 0} \frac{\partial h}{\partial g_{\mr{L}}}
\end{split}
\een
In addition if $h$ is time-dependent, $h(t;0)$ refers to the function evaluated at time $t$ in the
case where $g_{\mr{L}}=0$ and $g_{\mr{R}}=0$. So $\tilde{x}(t;0)$ is essentially the same object as
the operator $\hat{x}(t)$ for an arbitrary driving scalar field $f$ and where the system is
initially at the ground state.
Now let us consider the starting ansatz
\ben
|\Phi(t;g_{\mr{R}},g_{\mr{L}})\rangle = \exp\Big(\hat{T}+g_{\mr{R}}\hat{x}^N(t)+\tilde{x}(t)-\ui
[g_{\mr{R}}\phi^N(t)+\tilde{\phi}(t)]\Big)|0\rangle
\een
On the basis of the previous analysis, we derive from this wavefunction the following:
\ben
\begin{split}
|\Phi_{\mr{r}}(t)\rangle &=\lim_{g_{\mr{L}},g_{\mr{R}}\rightarrow 0}\frac{\partial}{\partial
g_{\mr{R}}}|\Phi(t;g_{\mr{R}})\rangle \\
&\approx |\Phi^{(0)}_{\mr{r}}(t)\rangle + |\Phi^{(1)}_{\mr{r}}(t)\rangle
\end{split}
\een
where
\ben
|\Phi^{(0)}_{\mr{r}}(t)\rangle =e^{\hat{T}-\ui E_0 t}[\hat{x}^N(t)-\ui \Delta \phi^N(t)]|0\rangle
\een
and $|\Phi^{(1)}_{\mr{r}}(t)\rangle$ is assigned as
\ben
|\Phi^{(1)}_{\mr{r}}(t)\rangle = [\tilde{x}_{\mr{r}}(t)-\ui \Delta\tilde{\phi}(t;0)\hat{x}^N(t)-\ui
\Delta\tilde{\phi}_{\mr{r}}(t)]e^{\hat{T}-\ui E_0 t}|0\rangle
\een
where $\Delta \phi^N$ is a relatively small residual term that would vanish in a formally exact
calculation.
In the above we neglected $\exp[\tilde{x}(t;0)]$ and few quadratic terms.
Similarly, the left ansatz reads
\ben
\begin{split}
\langle\Upsilon(t;g_{\mr{L}},g_{\mr{R}})| =& \langle 0|[\hat{L}_0+g_{\mr{L}}\hat{\lambda}^N(t)+\tilde{\lambda}(t)]
\exp\Big(-\hat{T}-g_{\mr{R}}\hat{x}^N(t)-\tilde{x}(t)\\&+\ui[E_0 t +g_{\mr{R}}\Delta \phi^N(t)+\Delta
\tilde{\phi}(t;g_{\mr{L}},g_{\mr{R}})]\Big)
\end{split}
\een
From this left ket the approximated state is derived:
\ben
\begin{split}
\langle \Upsilon_{\mr{l}}(t)| &=\lim_{g_{\mr{L}},g_{\mr{R}}\rightarrow 0}\frac{\partial}{\partial
g_{\mr{L}}}\langle \Upsilon (t;g_{\mr{L}},g_{\mr{R}})|\\
& \approx \langle \Upsilon^{(0)}_{\mr{l}}(t)| 
+ \langle \Upsilon^{(1)}_{\mr{l}}(t)|
\end{split} 
\een
where
\ben
\begin{split}
\langle \Upsilon^{(0)}_{\mr{l}}(t)| &=\langle 0|\hat{\lambda}^N(t)e^{-\hat{T}+\ui E_0 t}\\
\langle \Upsilon^{(1)}_{\mr{l}}(t)| &= \langle 0|[\tilde{\lambda}_{\mr{l}}(t)+\ui \Delta \tilde{\phi}(t;0)
\hat{\lambda}^N(t)+\ui\hat{L}_0\Delta\tilde{\phi}_{\mr{l}}(t)]e^{-\hat{T}+\ui E_0 t}
\end{split}
\een
The left and right response kets can be expanded in their respective eigenbasis
[$\tilde{x}(t)=\sum_I \tilde{c}_I(t)\hat{X}^I$, $\tilde{\lambda}(t)=\sum_I \tilde{d}_I(t)\hat{\Lambda}^I$], giving
\ben
|\Phi_{\mr{r}}^{(1)}(t)\rangle =  \Big\{\sum_I[\tilde{c}_{\mr{r},I}(t)- \delta_{NI}\ui\Delta \phi(t)] 
 \hat{X}^I-\ui\Delta \tilde{\phi}_{\mr{r}}(t)\Big\}e^{\hat{T}-\ui E_0 t}|0\rangle~,
\een 
and 
\ben
\langle \Upsilon_{\mr{l}}^{(1)}(t)| = \langle 0|  e^{-\hat{T}+\ui E_0 t}
\Big\{\sum_I\hat{\Lambda}^I[\tilde{d}_{\mr{l},I}(t)+
 \delta_{NI}\ui\Delta \phi(t)]+\hat{L}_0\ui\Delta\tilde{\phi}_{\mr{l}}(t)\Big\}
\een
where 
\ben
\begin{split}
\tilde{c}_{\mr{r},I}(t)&=\frac{\partial \tilde{c}_{I}}{\partial g_{\mr{R}}}\Big|_{g_{\mr{R}}=0}\\
\tilde{d}_{\mr{l},I}(t)&=\frac{\partial \tilde{d}_{I}}{\partial g_{\mr{L}}}\Big|_{g_{\mr{L}}=0,~g_{\mr{R}}=0}
\end{split}
\een
From the above equation we extract the following
approximated excited-state wave function $\langle \Upsilon^I| = \langle
0|\hat{\Lambda}^I\exp(-\hat{T})$, which leads to:
\ben
\langle \Upsilon^I|\Phi_{\mr{r}}^{(1)}(t)\rangle=[\tilde{c}_{\mr{r},I}(t)-
\delta_{NI}\ui\Delta \phi(t)]e^{-\ui E_0 t}\approx \mc{C}_I(t)
\een
\sloppy Analogously, using
$|\Phi^I\rangle = \hat{X}^I\exp(\hat{T})|0\rangle$ we see that $\langle
\Upsilon_{\mr{l}}^{(1)}(t)|\Phi^I\rangle = [\tilde{d}_{\mr{l},I}(t)+\delta_{NI}\ui\Delta\phi(t)]\exp(\ui E_0 t)
\approx \mc{C}_I^*(t)$.

The motion equations in this case follow from Eqs. (\ref{slr_x}) and (\ref{slr_l}):
\ben
\ui\partial_t \tilde{x}_{\mr{r},\mu}(t) = \Big\langle\hat{\tau}_{\mu}^{\dagger}\Big\{ 
  [\bar{H}_0,
  \tilde{x}_{\mr{r}}(t)]+[\bar{v}(t),\hat{x}^N(t)]+\Big[[\bar{H}_0,\hat{x}^N(t)],\tilde{x}(t;0)\Big]\Big\}\Big\rangle_0
\een
and 
\ben
-\ui\partial_t \tilde{\lambda}_{\mr{l},\mu}(t) = \Big\langle \tilde{\lambda}_{\mr{l}}(t)\bar{H}^0_{\tau,\mu}
 +\hat{\lambda}^N(t)\bar{v}_{\tau,\mu}(t)+\hat{\lambda}^N(t)[\bar{H}^0_{\tau,\mu},\tilde{x}(t;0)]\Big\rangle_0
\een
In the eigenbasis representation we then have that 
\ben
\begin{split}
\big(\ui\partial_t -\Omega_I\big)\tilde{c}_{\mr{r},I}(t)&=
e^{-\ui\Omega_N t}\Big\langle \hat{\Lambda}^I[\bar{v}(t),\hat{X}^N]+
\sum_J\hat{\Lambda}^I\Big[[\bar{H}_0,\hat{X}^N],\hat{X}^J\Big]c_J(t)
\Big\rangle_0\\
\big(-\ui\partial_t+\Omega_I\big)\tilde{d}_{\mr{l},I}(t)
&=e^{\ui\Omega_N t}\Big\langle \hat{\Lambda}^N[\bar{v}(t),\hat{X}^I]+
\sum_J\hat{\Lambda}^N\Big[[\bar{H}_0,\hat{X}^I],\hat{X}^J\Big]c_J(t)\Big\rangle_0
\end{split}
\een
Even though these two equations involve similar objects, they are different. 
Hence the left ($\tilde{d}_{\mr{l},I}$) and
right ($\tilde{c}_{\mr{r},I}$) amplitudes differ from one another.

The assignment deduced above can be applied to derive the excited-state transition elements in a
different way, by simply taking the functional derivatives and extracting the information from this.
Such feature can be seen if variation with respect to $f(t)$ are taken for coefficients such as
$\mc{C}_J$ and $\tilde{c}_{\mathrm{r},I}-\delta_{NI}\mathrm{i}\Delta\phi(t)$, where one would derive
an equation identical to Eq. (\ref{PsiIN}). Not only do quantum terms such as $\mc{C}_J$ lead to
transition matrix elements, but they are also an integral component in predicting the course of a
photo-stimulated physical process, or driven by other factors. Hence a connection between the CC
analogue is relevant to bridge electronic structure algorithms with photophysical models.

\section{General Evolution Equations}
\subsection{Extension of the SLR Framework}
We consider a more general propagation from an excited state, {\slshape i.e.}, 
$\hat{U}(t)|\Psi_N\rangle$, where $\hat{U}(t)=\mc{T}\exp[-\ui\int_0^t\ud s~\hat{H}(s)]$ ($\mc{T}$
being time-ordering super-operator), 
and extend our formalism beyond linear response; we refer to this as SR theory. First, we write
$
|\Psi(t;g_{\mr{R}})\rangle = \hat{U}(t) [|\Psi_0\rangle + g_{\mr{R}}
|\Psi_N\rangle]
$, and $\langle \Psi(t;g_{\mr{L}})| = [\langle \Psi_0| + g_{\mr{L}}
\langle\Psi_N|]\hat{U}^{\dagger}(t)$. We also define:
\ben
\langle A(t;g_{\mr{L}},g_{\mr{R}})\rangle = \langle \Psi(t;g_{\mr{L}})|\hat{A}|\Psi(t;g_{\mr{R}})\rangle
\een
Hence:
\ben\label{AHQM}
\lim_{g_{\mr{L}},g_{\mr{R}}\rightarrow 0}\Big[\frac{\partial }{\partial g_{\mr{L}}}+
\frac{\partial}{\partial g_{\mr{R}}}\Big]
\langle A(t;g_{\mr{L}},g_{\mr{R}})\rangle = \langle\Psi_N|\hat{A}^{\mr{H}}(t)|\Psi_0\rangle + \mr{c.c.}
\een
where $\hat{A}^{\mr{H}}(t)=\hat{U}^{\dagger}(t)\hat{A}\hat{U}(t)$. Note that the derivative above
has information about propagation of both the excited-state of interest {\slshape and} the ground
state of the system. In this case, a full normalization of the left and right initial
states ($\Psi(t;g_{\mr{L}}),~\Psi(t;g_{\mr{R}})$) is not
required as such normalization as no effect on the final result. 

The same notation applied before, Sections \ref{waveamp} and \ref{slr_theory}, is used in this
section to derive the SR equations. We start with the set shown in Eq. (\ref{slr_ansatz}) and inserting these in Eqs.
(\ref{x_eq}-\ref{phi_eq}), where no further assumptions are taken. So the response operators
$\tilde{x}(t)$, $\tilde{\lambda}(t)$, and the phase $\tilde{\phi}(t)$ are now valid for arbitrary
strengths of the perturbation. It is important to bear in mind that the operators $\hat{\lambda}(t)$ and
$\hat{x}(t)$, when the system does not initiate completely from a ground-state configuration, are functions
of the numbers $g_{\mr{L}}$ and $g_{\mr{R}}$, allowing us to compute variations of these operators
with respect to such parameters at any time $t$, including $t=0$, leading to the equations discussed
below.

In the present case, the expectation value reads $\langle A(t;g_{\mr{L}},
g_{\mr{R}})\rangle = \langle
\Upsilon(t;g_{\mr{L}},g_{\mr{R}})|\hat{A}|\Phi(t;g_{\mr{L}},g_{\mr{R}})\rangle$, so it satisfies:
\ben\label{AH}
\begin{split}
\lim_{g_{\mr{L}},g_{\mr{R}}\rightarrow 0}&\Big[\frac{\partial }{\partial
g_{\mr{L}}}+\frac{\partial}{\partial g_{\mr{R}}}\Big]
\langle A(t;g_{\mr{L}},g_{\mr{R}})\rangle = \langle
\hat{\lambda}_{\mr{l}}(t)
e^{-\tilde{x}(t;0)}\bar{A}e^{+\tilde{x}(t;0)}\rangle_0\\
&+\langle [\hat{L}_0+\hat{\lambda}(t;0)]
e^{-\tilde{x}(t;0)}[\bar{A},\hat{x}_{\mr{r}}(t)]e^{+\tilde{x}(t;0)}
+\hat{\lambda}_{\mr{r}}(t)e^{-\tilde{x}(t;0)}\bar{A}e^{+\tilde{x}(t;0)}\rangle_0
\end{split}
\een
Because we distinguish the parameters $g_{\mr{L}}$ and $g_{\mr{R}}$, we assign the term as
$\langle\Psi_N|\hat{A}^{\mr{H}}(t)|\Psi_0\rangle$ as $\langle \hat{\lambda}_{\mr{l}}(t)
e^{-\tilde{x}(t;0)}\bar{A}e^{+\tilde{x}(t;0)}\rangle_0$, and the other quantity containing the
right-handed derivatives as $\langle\Psi_0|\hat{A}^{\mr{H}}(t)|\Psi_N\rangle$ (in the numerical
calculations shown in the next section we found they are visually identical, but in more practical contexts
they are not expected to be so).
Where the general equations of motion are 
\ben
\begin{split}
\ui\partial_t x_{\mr{r},\mu}(t) &= \langle \hat{\tau}^{\dagger}_{\mu}
e^{-\tilde{x}(t;0)}[\bar{H}(t),\hat{x}_{\mr{r}}(t)]e^{+\tilde{x}(t;0)}\rangle_0\\
-\ui\partial_t \lambda_{\mr{l},\mu}(t)&=
\langle \hat{\lambda}_{\mr{l}}(t)
e^{-\tilde{x}(t;0)}[\bar{H}(t),\hat{\tau}_{\mu}]e^{+\tilde{x}(t;0)}\rangle_0\\
-\ui\partial_t \lambda_{\mr{r},\mu}(t)&=
\Big\langle [\hat{L}_0+\tilde{\lambda}(t;0)]e^{-\tilde{x}(t;0)}[\bar{H}_{\tau,\mu}(t),
\hat{x}_{\mr{r}}(t)]e^{+\tilde{x}(t;0)}
+\hat{\lambda}_{\mr{r}}(t)e^{-\tilde{x}(t;0)}\bar{H}_{\tau,\mu}(t)e^{+\tilde{x}(t;0)}\Big\rangle_0\\
\Delta\phi_{\mr{r}}(t)&=\int_0^t\ud s~\Big\langle  
[\hat{L}_0+\tilde{\lambda}(s;0)]
e^{-\tilde{x}(s;0)}[\bar{H}(s),\hat{x}_{\mr{r}}(s)]e^{+\tilde{x}(s;0)}
+\hat{\lambda}_{\mr{r}}(s)e^{-\tilde{x}(s;0)}\bar{H}(s)e^{+\tilde{x}(s;0)} \Big\rangle_0\\
\Delta\phi_{\mr{l}}(t)&=\int_0^t\ud s~\Big\langle 
\hat{\lambda}_{\mr{l}}(s)
e^{-\tilde{x}(s;0)}\bar{H}(s)e^{+\tilde{x}(s;0)} \Big\rangle_0
\end{split}
\een
recall that $\bar{H}_{\tau,\mu}(t)=[\bar{H}(t),\hat{\tau}_{\mu}]$ refers to the full Hamiltonian.
The operators $\tilde{\lambda}(t;0)$ and $\tilde{x}(t;0)$ refer to the solution of Eqs. (\ref{x_eq})
and (\ref{l_eq}) in the case where the system is initially at the ground-state, so
$\tilde{x}(t=0;0)=\tilde{\lambda}(t=0;0)=0$.
Also note the equations for the phases correspond to taking the derivatives of the quantity $\Delta
\phi(t)$, not $\tilde{\phi}$.
The initial conditions of the (de)excitation amplitudes are
$\hat{\lambda}_{\mr{l}}(t=0)=\hat{\lambda}^N(t=0)$, $\hat{x}_{\mr{r}}(t=0)=\hat{x}^N(t=0)$, 
and for $\hat{\lambda}_{\mr{r}}$ we have:
\ben
\hat{\lambda}_{\mr{r}}(t=0)=-\sum_I \frac{F^{NI}}{\Omega_N+\Omega_I}\hat{\Lambda}^I
\een
These initial conditions ensure that at the initial time, the transition moment $\langle
\Psi_N|\hat{A}|\Psi_0\rangle$ is consistent with the standard CC linear response result.
On the other hand, for the energy $\langle E(t;g_{\mr{L}}, g_{\mr{R}})\rangle=\langle
\Upsilon(t;g_{\mr{L}},g_{\mr{R}})|\hat{H}(t)|\Phi(t;g_{\mr{L}},g_{\mr{R}})\rangle $,
using the phases above we obtain
\ben
\lim_{g_{\mr{L}},g_{\mr{R}}\rightarrow 0}\Big[\frac{\partial }{\partial
g_{\mr{L}}}+\frac{\partial}{\partial g_{\mr{R}}}\Big]
\langle E(t;g_{\mr{L}},g_{\mr{R}})\rangle = \partial_t [\Delta\tilde{\phi}_{\mr{l}}(t)+\Delta\tilde{\phi}_{\mr{r}}(t)]
\een
This result provides a connection between the phase response and the energy evolution.

\sloppy Equation (\ref{AH}) is advantageous as it provides linked expressions for quantities
such as $\langle \Psi_N|\hat{A}^{\mr{H}}|\Psi_0\rangle$, in which
the ground- and excited-state propagations are present together. 
The resolution of the identity can be inserted on both sides of the operator
$\hat{A}$, which gives for instance:
$
\langle \Psi_N|\hat{A}^{\mr{H}}(t)|\Psi_0\rangle = \sum_{IJ} \langle
\Psi_N|\hat{U}^{\dagger}(t)|\Psi_I\rangle A_{IJ} \langle \Psi_J|\hat{U}(t)|\Psi_0\rangle
$
where $A_{IJ} = \langle \Psi_I|\hat{A}|\Psi_J\rangle$. Therefore the expression above has
contributions from the solutions to the excited- and ground-state problems. The excited-state
component can be extracted through a frequency space analysis, or a related technique. 

\sloppy Alternatively, a single resolution operation can be applied, giving 
\ben
\langle \Psi_N|\hat{A}^{\mr{H}}(t)|\Psi_0\rangle=\sum_J \mc{C}^*_J(t) \langle \Psi_J|\hat{A}\hat{U}(t)|\Psi_0\rangle
\een
($\mc{C}^*_J(t)=\langle \Psi_N|\hat{U}(t)|\Psi_J\rangle$).  If this idea is applied to the first term on
the right hand side of Eq. (\ref{AH}), we obtain the two elements:
$\delta_{NJ}\exp(\ui\Omega_Nt)+\tilde{d}_{\mr{l},J}(t)$ and $\langle \hat{\Lambda}^J
\exp[-\tilde{x}(t;0)]\bar{A}\exp[+\tilde{x}(t;0)]\rangle_0$.  These resemble in appearance their
parent linear (quantum mechanical) counter-parts, from Eq. (\ref{AHQM}). Hence it is plausible to approximate $C^*_J(t)$
using $\tilde{d}_{\mr{l},J}+\ui\delta_{JN}\Delta\phi(t)$, where $\tilde{d}_{\mr{l},J}(t)=\langle
\tilde{\lambda}_{\mr{l}}(t)\hat{X}^J\rangle_0$. Although the right-handed contribution is more interconnected
than the left one, it may be associated approximately to the term $\langle
\Psi_0|\hat{A}^{\mr{H}}(t)|\Psi_N\rangle$. In the next section we use a numerical model to discuss
the right-handed expression for $C_J(t)$.

Although the assignment above might serve useful for interpretation and for quantitative analysis,
it could result in more rigorous formulas a direct comparison in frequency space based on the
specific form of the perturbation used. A robust determination of the TD element $\langle
\Psi_I|\hat{U}(t)|\Psi_N\rangle$ for a manyfold of $N, I$ states in turn provides a non-symmetric
representation of the operator $\hat{U}(t)$ and by extension a propagator for general initial states
of the form $|\Psi(t=0)\rangle=\sum_J C_{J,0}|\Psi_J\rangle$. This supposes that the propagator is
represented in the eigenbasis of the Hamiltonian. It is possible, however, to change the basis
representing the operators, such as that corresponding to the bare single orbital excitations,
characterized by the indices $\mu$ and $\nu$. The choice is largely dependent on the potential
numerical approach of interest. We pursue the excited-state energy picture because of its connection
to physical models, where a state-by-state perspective becomes convenient and leads to the
calculation and understanding of optical and/or magnetic spectra.

\subsection{Propagation from an Arbitrary Initial State}
It is possible to obtain the time-evolution of an observable average where the system is an initial
state described by a linear combination of eigenstates. We thus denote:
$
|\Psi_{\mr{R}}\rangle = \mc{N}^{-1/2}[|\Psi_0\rangle+g_{\mr{R}}|\Psi(0)\rangle]$, 
and $\langle \Psi_{\mr{L}}| = \mc{N}^{-1/2}[\langle\Psi_0|+g_{\mr{L}}\langle \Psi(0)|]$, where
$\mc{N}$ is the normalization factor 
\ben
\mc{N} = 1+g_{\mr{L}}S+g_{\mr{R}}S^*+g_{\mr{L}}g_{\mr{R}}
\een
and $S$ is the overlap between the ground-state and the initial wavefunctions: $S=\langle
\Psi(0)|\Psi_0\rangle$.
The initial wave function reads:
\ben
|\Psi(0)\rangle=\sum_{N}C_N|\Psi_N\rangle
\een
The set $\{C_N\}$ represents normalized complex-valued coefficients ($\sum_N |C_N|^2=1$). Contrary to the case
of expressing $\langle \Psi_N|\hat{A}^{\mr{H}}(t)|\Psi_0\rangle$, in this instance the normalization
function $\mc{N}$ is of crucial relevance.

To obtain the element $\langle \Psi(0)|\hat{A}^{\mr{H}}|\Psi(0)\rangle$ we apply the following limit
to the mixed second-degree derivative, which gives:
\ben
\lim_{g_{\mr{L}},g_{\mr{R}}\rightarrow 0}\frac{\partial^2}{\partial g_{\mr{L}}\partial g_{\mr{R}}}
\langle \Psi_{\mr{L}}|\hat{A}^{\mr{H}}(t)|\Psi_{\mr{R}}\rangle = 
\langle \Psi(0)|\hat{A}^{\mr{H}}(t)|\Psi(0)\rangle-\langle\Psi_0|\hat{A}^{\mr{H}}(t)|\Psi_0\rangle-
\mc{I}(t)
\een
where 
\ben
\mc{I}(t) = [S\langle\Psi(0)|\hat{A}^{\mr{H}}(t)|\Psi_0\rangle+\mr{c.c.}]-2S\times
S^*\langle\Psi_0|\hat{A}^{\mr{H}}(t)|\Psi_0\rangle
\een
In the standard picture the element $\langle\Psi(0)|\hat{A}^{\mr{H}}(t)|\Psi(0)\rangle$ is equivalent to $\langle\Psi(t)|\hat{A}|\Psi(t)\rangle$,
with $|\Psi(t)\rangle=\hat{U}(t)|\Psi(0)\rangle$.
In this case we then use a different initial condition for the cluster operators, so 
$\hat{\lambda}_{\mr{l}}(t=0)=\sum_M C^*_M\hat{\Lambda}^M$, and $\hat{x}_{\mr{r}}(t=0)=\sum_N
C_N\hat{X}^N$. The superposition of operators does not translate into a superposition of symmetrized
wavefunction, but instead it ensures that at the end of the calculation one obtains
$\langle \Psi(0)|\hat{A}^{\mr{H}}|\Psi(0)\rangle=\sum_{M,N}C_M^*C_N\langle
\Psi_M|\hat{A}^{\mr{H}}|\Psi_N\rangle$. 

With the initial conditions defined we derive the
expression:
\ben
\begin{split}
\lim_{g_{\mr{L}},g_{\mr{R}}\rightarrow 0}\frac{\partial^2}{\partial g_{\mr{L}}\partial g_{\mr{R}}}
\langle &\Upsilon(t;g_{\mr{L}},g_{\mr{R}})|\hat{A}|\Phi(t;g_{\mr{L}},g_{\mr{R}}\rangle
=\\&\langle \hat{\lambda}_{\mr{l}}(t)e^{-\tilde{x}(t;0)}[\bar{A},\hat{x}_{\mr{r}}(t)]
e^{+\tilde{x}(t;0)}+\hat{\lambda}_{\mr{l,r}}(t)e^{-\tilde{x}(t;0)}\bar{A}e^{+\tilde{x}(t;0)}\rangle_0
\end{split}
\een
where $\hat{\lambda}_{\mr{l,r}}$ is the mixed derivative ($\partial^2\hat{\lambda}/\partial
g_{\mr{L}}\partial g_{\mr{R}}$) with respect to $g_{\mr{L}}$ and
$g_{\mr{R}}$ evaluated at $g_{\mr{L}}=g_{\mr{R}}=0$, and it follows the motion equation:
\ben
-\ui\partial_t \lambda_{\mr{l,r},\mu}(t)=\langle \hat{\lambda}_{\mr{l,r}}(t)
e^{-\tilde{x}(t;0)}\bar{H}_{\tau,\mu}(t)e^{+\tilde{x}(t;0)}
+\hat{\lambda}_{\mr{l}}(t)
e^{-\tilde{x}(t;0)}[\bar{H}_{\tau,\mu}(t),\hat{x}_{\mr{r}}(t)]e^{+\tilde{x}(t;0)}
\rangle_0
\een
in which 
\ben
\hat{\lambda}_{\mr{l,r}}(t=0) = \sum_J Y_J\hat{\Lambda}^J
\een
and
\ben\label{Yj}
Y_J =
\sum_{M,N}C_M^*(0)C_N(0)\frac{\langle\hat{\Lambda}^M\big[[\bar{H}_0,\hat{X}^N],\hat{X}^J\big]\rangle_0}
{\Omega_M-\Omega_J-\Omega_N}
\een
This initial condition guarantees that at the initial propagation time the element
$\langle\Psi(0)|\hat{A}^{\mr{H}}(0)|\Psi(0)\rangle$ is consistent with quadratic response theory.

Using the standard TD CC equations for ground-state propagation ($\langle
\Psi_0|\hat{A}^{\mr{H}}(t)|\Psi_0\rangle$, we find the relation:
\ben\label{ANN}
\begin{split}
\langle\Psi(0)|&\hat{A}^{\mr{H}}(t)|\Psi(0)\rangle
= \langle [\hat{L}_0+\tilde{\lambda}(t;0)] e^{-\tilde{x}(t;0)}\bar{A}
e^{+\tilde{x}(t;0)}\rangle_0\\&+\langle\hat{\lambda}_{\mr{l}}(t)e^{-\tilde{x}(t;0)}[\bar{A},\hat{x}_{\mr{r}}(t)]
e^{+\tilde{x}(t;0)}+\hat{\lambda}_{\mr{l,r}}(t)e^{-\tilde{x}(t;0)}\bar{A}e^{+\tilde{x}(t;0)}\rangle_0
+\mc{I}(t)
\end{split}
\een
If the ground state wavefunction $\Psi_0$ is orthogonal to the initial state then $\mc{I}=0$,
otherwise this term, $\mc{I}(t)$, can be computed using the Eqs. (\ref{eq_eleven}) and
(\ref{eq_lambda_t}). Our method is
also applicable to obtain an element such that $\langle \Psi_J|\hat{A}^{\mr{H}}(t)|\Psi_I\rangle$.
This only requires changing the initial conditions of the left and right cluster operators, and a
simple adaptation of Eq. (\ref{Yj}) where $C_M^*(0)$ and $C_N(0)$ are replaced by $\delta_{JM}$ and
$\delta_{IN}$, correspondingly, and the same applies to the initial conditions.  In fact one can
analyze propagating the wavefunctions $\langle \Psi_\mr{L}|
= \langle \Psi_0|+g_{\mr{L}}\langle \Psi_J|$ and $|\Psi_R\rangle=
|\Psi_0\rangle+g_{\mr{R}}|\Psi_I\rangle$ and conclude that our formalism gives the element $\langle
\Psi_J|\hat{A}^{\mr{H}}(t)|\Psi_I\rangle$ in terms of the equations shown above, with the
mentioned required adaptations. This would in turn justify the initial conditions for cluster
operators we applied to obtain the general evolution of a quantum mechanical observable under an
arbitrary initial state, $\langle \Psi(0)|\hat{A}^{\mr{H}}(t)|\Psi(0)\rangle$.

\section{Numerical Illustration}
\begin{figure}[htb!]
\centering
\includegraphics[scale=0.65]{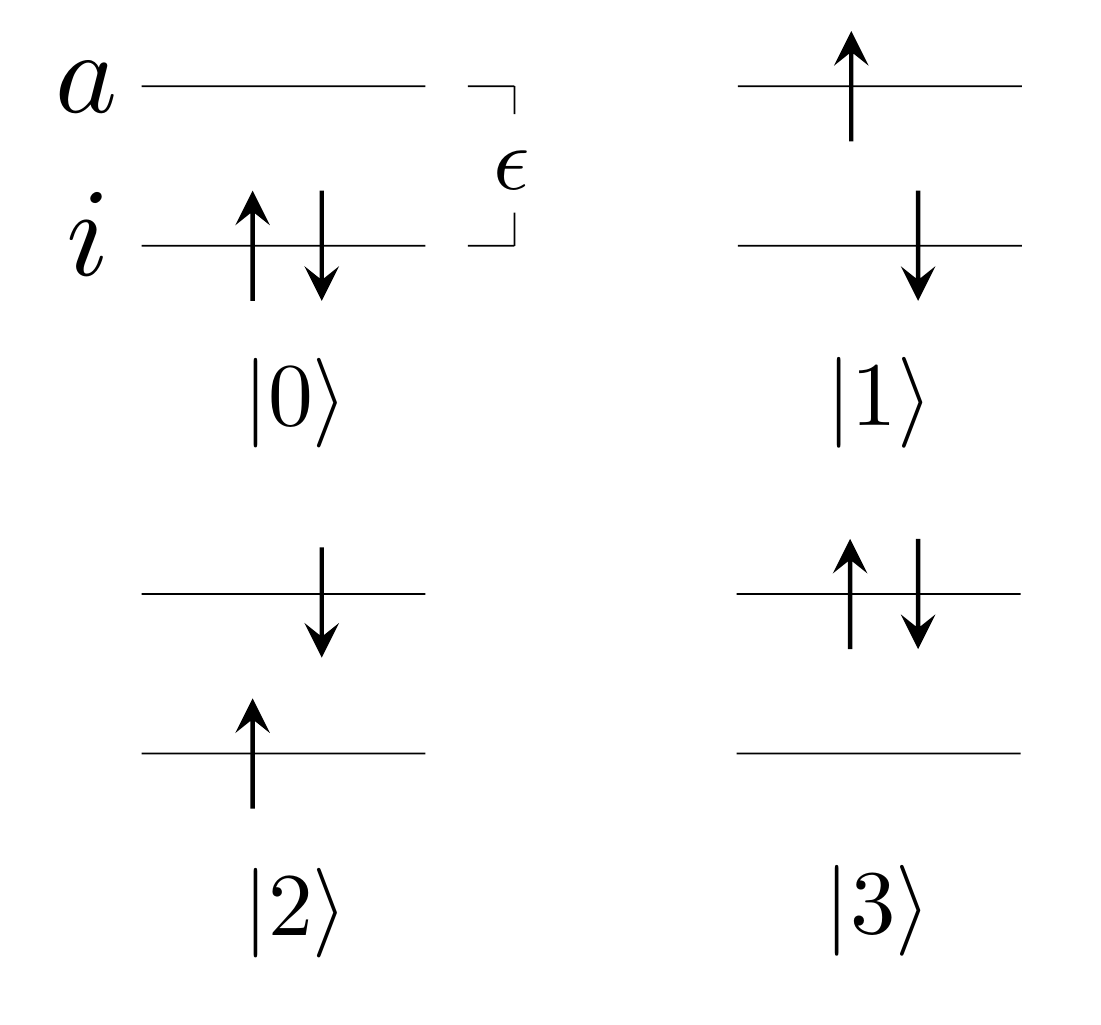}
\caption{Sketch of the two-level system considered for the numerical illustration.}\label{twolevel}
\end{figure}

Here we examine the application of our generalized SR method to a two-electron-two-level system, where we examine
in total four levels. It is studied here how the quantum system evolves under the presence of an
external TD driving field that is strong.
The Hamiltonian of the system is:
\ben
\label{Heq}
\hat{H}(t) = \sum_{\sigma}\epsilon \hat{a}^{\dagger}_{\sigma}\hat{a}_{\sigma} + \sum_{\sigma}
b[\hat{\tau}_{\sigma}+\hat{\tau}_{\sigma}^{\dagger}] +
w[\hat{\tau}_{\uparrow}\hat{\tau}_{\downarrow}+\hat{\tau}_{\downarrow}^{\dagger}\hat{\tau}^{\dagger}_{\uparrow}]
+\hat{v}(t)
\een
We denote the occupied level as $i$ and the unoccupied one as $a$, so
$\hat{\tau}_{\sigma}=\hat{a}^{\dagger}_{\sigma}\hat{i}_{\sigma}$. The external driving term reads
$\hat{v}(t)=-f(t)\mu_0\sum_{\sigma}[\hat{\tau}_{\sigma}+\hat{\tau}_{\sigma}^{\dagger}]$. The
function $f(t)$ describes a Gaussian pulse $f(t)=f_0\exp[-(t-t_0)/2\sigma_0^2]$.  In our simulation
we take $\epsilon$ as 1 eV, $b$ and $w$ as 0.25 eV, $\mu_0 = 0.5$ au, $f_0\mu_0$ as 1 eV (so
$f_0\approx 0.0735$ au, which is approximately $3.8\times 10^{10}$ V/m),
$\sigma_0=5~\mr{fs}$, and $t_0=2.5\times\sigma_0$. This corresponds to the applying a strong
pulse to the system.

In Fig. \ref{twolevel} we show the four mentioned quantum levels, which form the linear space we
consider: the ground-state
configuration $|0\rangle$, two separate single-electron promoted states, $|1\rangle$ and
$|2\rangle$, respectively, and the a doubly excited configuration, $|3\rangle$. All our
wavefunctions are constrained to the space $\mc{L}$ spanned by the set of mentioned states,
$\{|0\rangle,~|1\rangle,~|2\rangle,~|3\rangle\}$. We then translate all the required operators, such
as the Hamiltonian and the cluster operators, into matrix form over the basis shown in Fig.
\ref{twolevel}; this allows us to perform all the operations numerically. The diagonalization of the
Hamiltonian matrix reveals a considerable mixing between the states in the generation of the
eigenvectors; such mixing ensures that our model is {\slshape non-trivial}, which leads to the
characteristic asymmetries of non-Hermitian CC approaches, discussed below. The eigenvectors of the Hamiltonian matrix are referred to as
$\Psi_0,~\Psi_1,~\Psi_2$, and $\Psi_3$, where $\mb{H}_0|\Psi_J\rangle = E_J|\Psi_J\rangle$ (for
$J=0,1,2,3$).
The ground state is composed approximately of 92 \% $|0\rangle$ and 8 \% of the
singles configurations. The first excited state contains 13 \% of the doubles configuration
$|3\rangle$, 6 \% of $|0\rangle$, and the rest is equal mix of singles. The second excited state
is a triplet state with equal amounts of the $|1\rangle$ and $|2\rangle$ states. And the third
excited state is dominated by the doubles state $|3\rangle$ with a weight of 87 \%, the combined
states $|1\rangle$ and $|2\rangle$ give a weight of 10 \%, and the rest corresponds to $|0\rangle$.
Therefore there is considerable interaction by the configurations that we selected, Fig. \ref{twolevel}. 
The standard unitary operations based on the operator
$\hat{U}(t)$ were performed using a simple midpoint rule, where we discretize the whole time
interval as a grid and propagate step by step using $|\Psi(t+\delta t)\rangle \approx \exp(-\ui
\hat{H}(t+\delta t/2)\delta t)|\Psi(t)\rangle$.  For the TD CC equations we use the second order
Runge-Kutta methodology, over the same grid for the unitary propagation, which consists of sixty
thousand points.

\begin{figure}[htb!]
\centering
\subfigure[]{\includegraphics[scale=0.6]{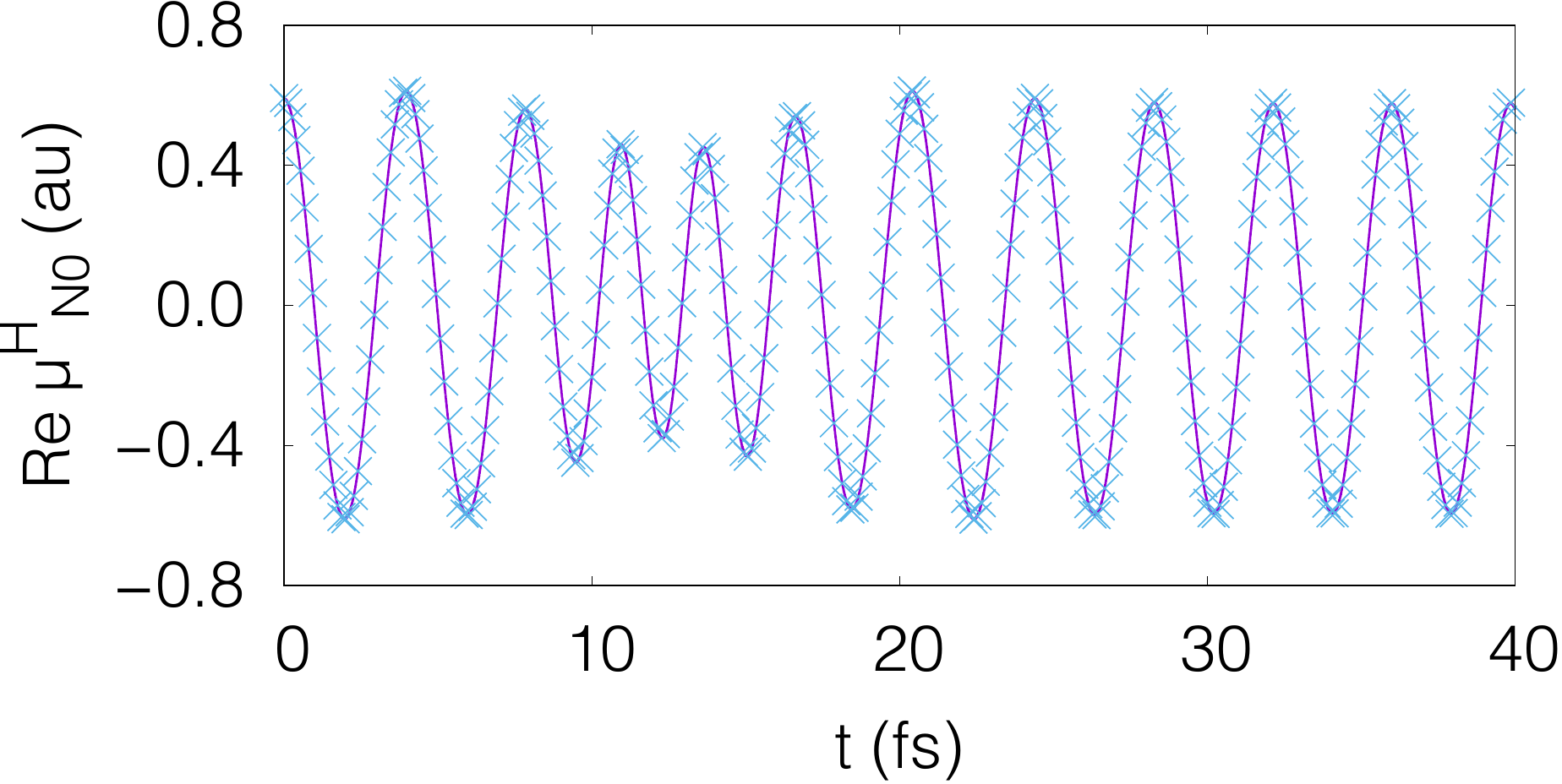}}
\subfigure[]{\includegraphics[scale=0.6]{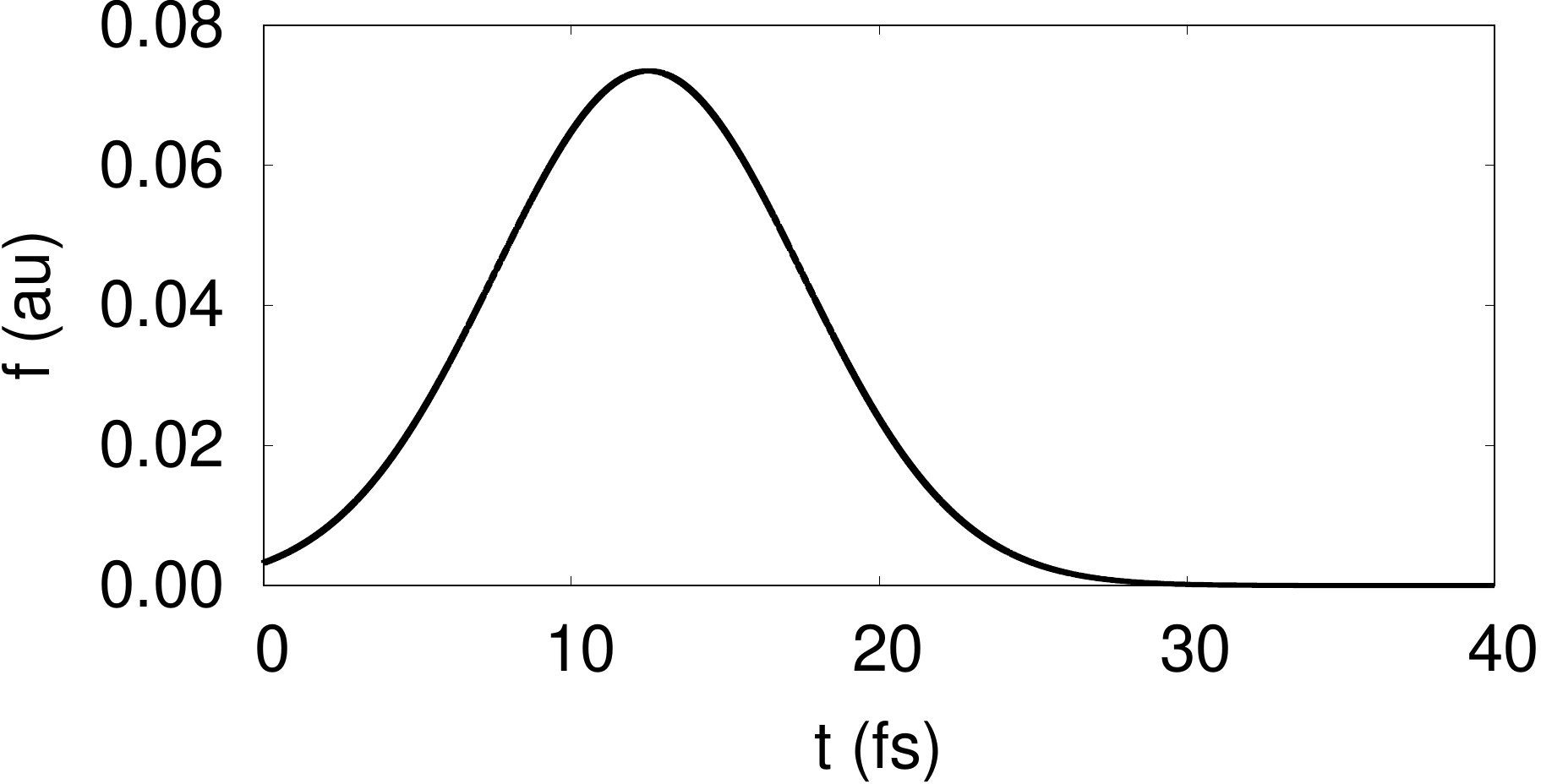}}
\caption{a) Computed element $\langle \Psi_N|\hat{A}^{\mr{H}}(t)|\Psi_0\rangle$
($\hat{A}=\hat{\mu}$) for $N=1$ in the interval of time
between $0$ and $40$ fs. The purple line corresponds to our second response calculations, whereas
the blue ``X'' symbols to sample points of full standard propagation, from computing $\hat{U}(t)$ in matrix
representation and applying it to the states $\Psi_N$ and $\Psi_0$, which are eigenfunctions of the
Hamiltonian matrix derived from Eq. (\ref{Heq}).}\label{AHH}
\end{figure}

Let us begin considering the computation of the element $\langle \Psi_N
|\hat{A}^{\mr{H}}(t)|\Psi_0\rangle$, where $\hat{A}$ corresponds to the dipole operator, which we
take in this work as $\hat{\mu}= \mu_0(\hat{\tau}_{\uparrow}+\hat{\tau}_{\downarrow}+\mr{H.c})$, and
denote $\langle \Psi_N |\hat{\mu}^{\mr{H}}(t)|\Psi_0\rangle$ as $\mu_{N0}^{\mr{H}}$.  The term
$\langle \Psi_N |\hat{A}^{\mr{H}}(t)|\Psi_0\rangle$ is an important quantity because in the
Heisenberg representation, for a general initial state that is a linear combination of other
eigen-states, a quantity of this kind is required.
For this reason we propose a model for this type of object because it would be needed for a
propagation from an initial state that includes a portion of the ground state. We take $N=1$, so our
simulation is based on propagating with the SR equations both the ground-state and the
excited-state. $\Psi_1$ is a singlet excited state of the system. Our basis
misses the two paramagnetic states in which the second level is occupied
with electron with the same $z$-spin as the electron in the first level. However, focus on singlet states. Fig. \ref{AHH}.a shows the time-dependency of the real part
of this object (its imaginary component behaves in a similar fashion) and Fig. \ref{AHH}.b the
shape of the pulse applied to the system. As expected, given that TD CC theory is robust if the
cluster operators cover all excitation orders, the SR theory and the standard unitary solution yield
visually identical results. Both the SR theory left and right expressions for the matrix element in the
Heisenberg representation offer the same results. This would not hold if the cluster operators are
truncated, which happens in practice; in that case the expressions may differ.

\begin{figure}[htb!]
\centering
\includegraphics[scale=0.65]{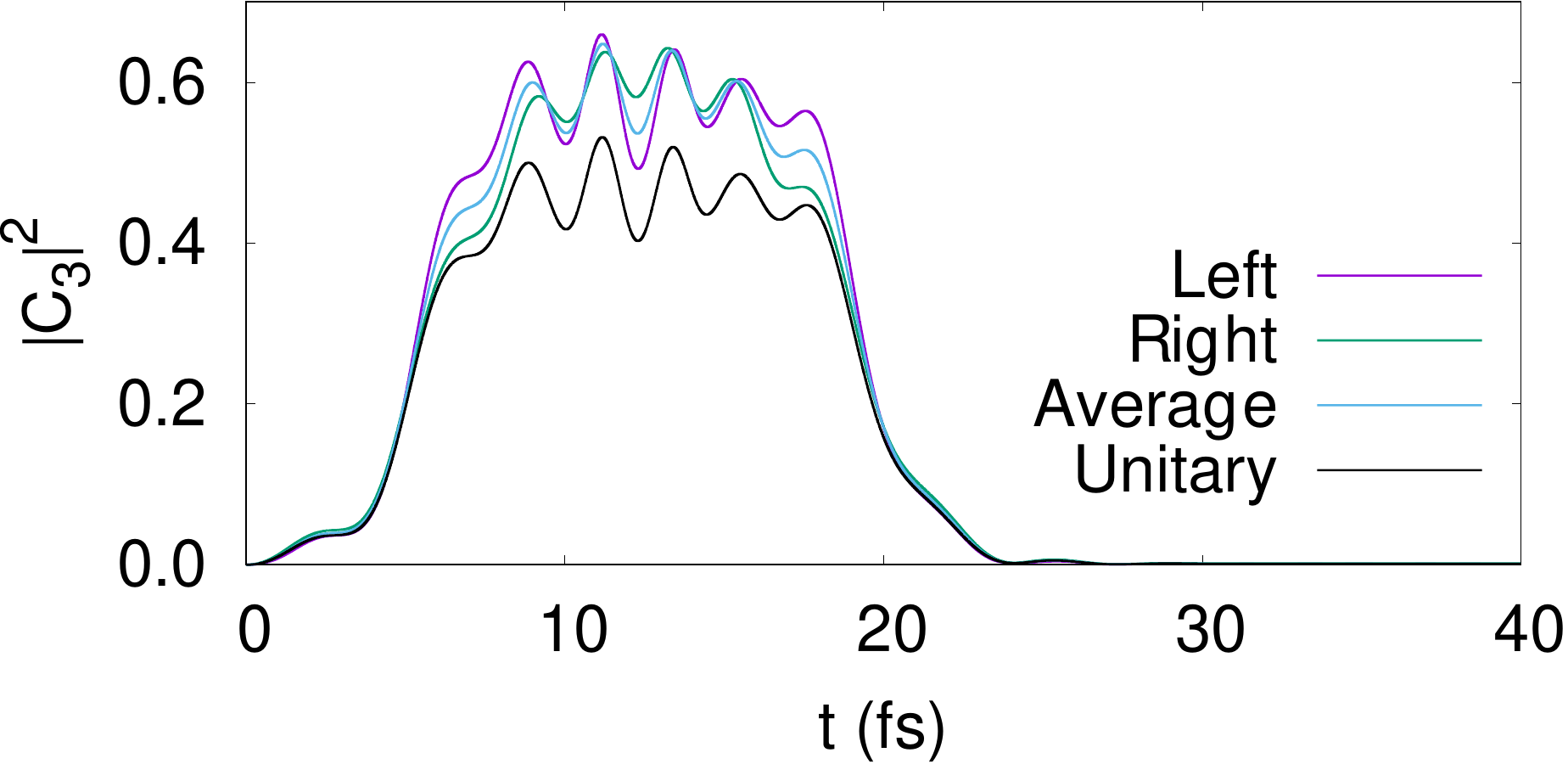}
\caption{Probability of finding the system in the doubly excited state $\Psi_3$,
when the quantum system evolves from state $\Psi_1$ and in the presence of the pulse showed in Fig.
\ref{AHH}. The black line is obtained from the unitary propagation, the purple line from
$\tilde{d}_{\mr{l},3}^*(t)$, the
green one from $c'_3(t)$, and blue line corresponds to the average
$1/2[c_3'(t)+\tilde{d}_{\mr{l},3}^*(t)]$.}\label{c3square}
\end{figure}

If this two-electron quantum system initiated evolution from the first excited state, then one can
ask about the probability of finding the system in the third excited state at some given time. Such
probability is determined by the squared modulus of the coefficient $C_3(t)=\langle
\Psi_3|\hat{U}(t)|\Psi_1\rangle$. This coefficient is approximated as $\tilde{d}_{\mr{l},J}^*(t)$ ($J=3$), which is discussed
in the previous section. For the right-handed contribution, we noted that the coefficient $c_J(t)$
often underestimates $C_3$ by a significant margin. As an alternative to this, we compute $c_J'(t)=
\langle \hat{\Lambda}^J\hat{x}_{\mr{r}}(t)\exp(\tilde{x}(t;0))\rangle_0/\lVert
\hat{x}_{\mr{r}}(t)\exp(\tilde{x}(t;0))|0\rangle\rVert$, and denote that as our right-handed
estimator. Computing the norm of
$\hat{x}_{\mr{r}}(t)\exp(\tilde{x}(t;0))|0\rangle$ is not practical for molecular systems due to the
need for Hermitian conjugation, but in this case the small size of the system allows for its computation.
We refer to $c_J'(t)$ as the right-handed approximation to the standard coefficient 
$\langle \Psi_J|\hat{U}(t)|\Psi_1\rangle$. Fig. \ref{c3square} shows the result of this procedure.
As discussed before, at short times our assignment holds, but as the pulse action becomes more
significant some deviations are present. Part of the reason for such behavior is the non-negligible
cluster amplitudes associated to the operator $\hat{T}$. We noticed that upon reducing the parameters $b$
and $w$ to about 0.1 eV, the agreement with respect $C_3$ is quite improved, especially for the
averaged value $c_{\mr{avg},3}(t) = 1/2\times [c_3'(t)+\tilde{d}_{\mr{l},3}^*(t)]$, but we believe it important to
emphasize potential deviations over closer agreements. 

\begin{figure}[htb!]
\centering
\includegraphics[scale=0.65]{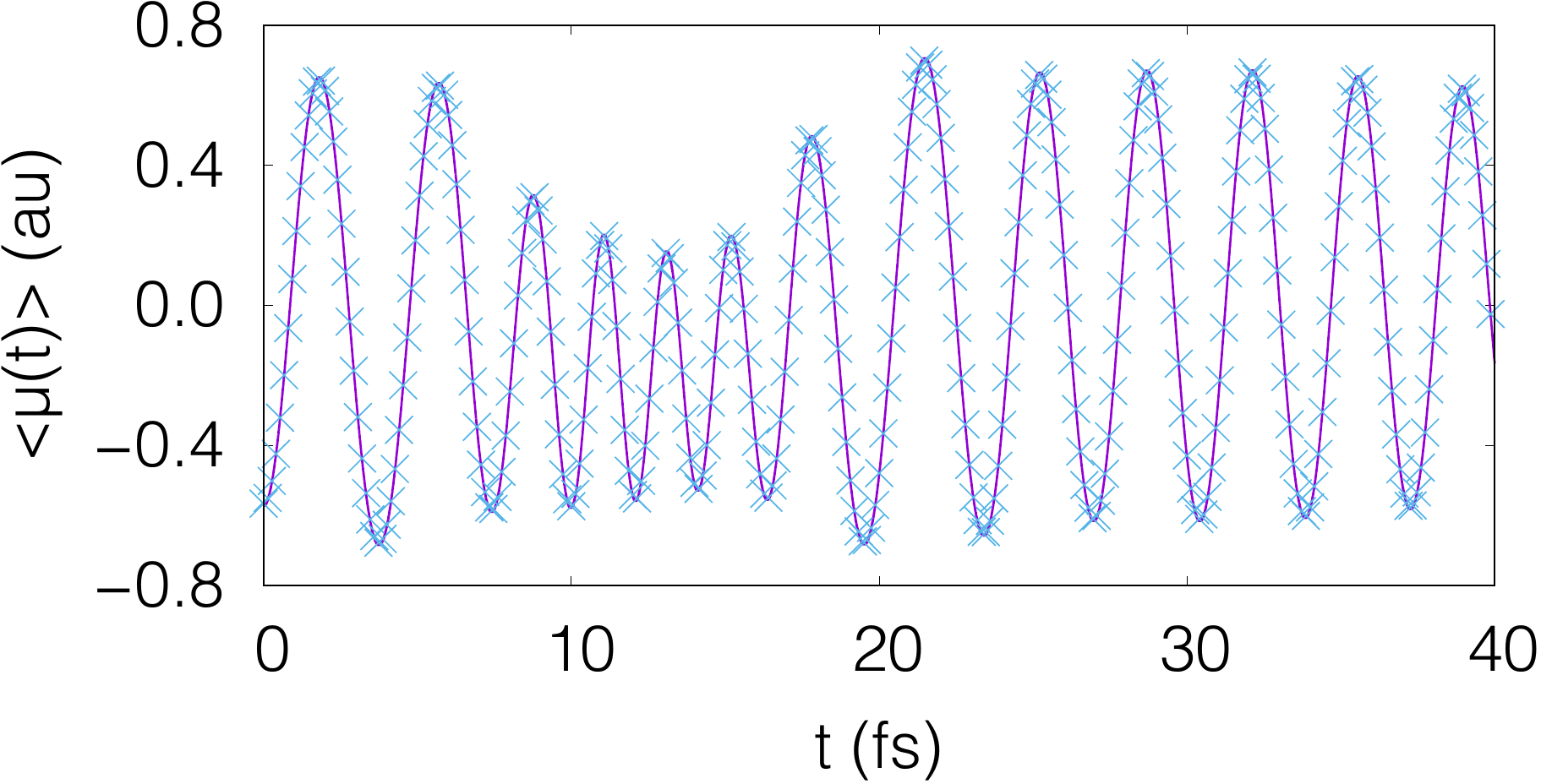}
\caption{Comparison between SR theory and unitary propagation for the calculation of the
time-dependent dipole of the system $\langle
\mu(t)\rangle=\langle\Psi(0)|\hat{U}^{\dagger}(t)\hat{\mu}\hat{U}(t)|\Psi(0)\rangle$, where
$|\Psi(0)\rangle$ is a linear combination of the states $|\Psi_1\rangle$ and $|\Psi_3\rangle$:
$|\Psi(0)\rangle = \sqrt{3/4} |\Psi_1\rangle+\sqrt{1/4} |\Psi_3\rangle$. Purple line: SR theory,
blue ``X'' symbols: samples from the unitary propagation.}\label{APsi}
\end{figure}

Now we show the application of the SR theory to compute the evolution of an observable such as the
dipole in the case where the system does not initiate at the ground state, but at a linear
combination of two excited states. We then choose as the initial state: 
\ben
|\Psi(0)\rangle = \sqrt{3/4} |\Psi_1\rangle+\sqrt{1/4}|\Psi_3\rangle
\een 
where the wavefunctions $|\Psi_1\rangle$ and $|\Psi_3\rangle$, in the basis shown in Fig.
\ref{twolevel}, correspond to the first and third excited states obtained from the diagonalization
of the unperturbed system Hamiltonian matrix. As in the case for calculating $\langle
\Psi_N|\hat{\mu}^{\mr{H}}(t)|\Psi_0\rangle$, the SR expression, Eq. (\ref{ANN}) with
$\hat{A}=\hat{\mu}$, for $\langle \mu(t)\rangle=\langle\Psi(t)|\hat{\mu}|\Psi(t)\rangle$ (where
$\hat{U}(t)|\Psi(0)\rangle=|\Psi(t)\rangle$) is fully consistent with respect to the unitary
propagation, Fig. \ref{APsi}, confirming the possibility of propagating an observable based on
a general initial state.

\begin{figure}[htb!]
\centering
\includegraphics[scale=0.45]{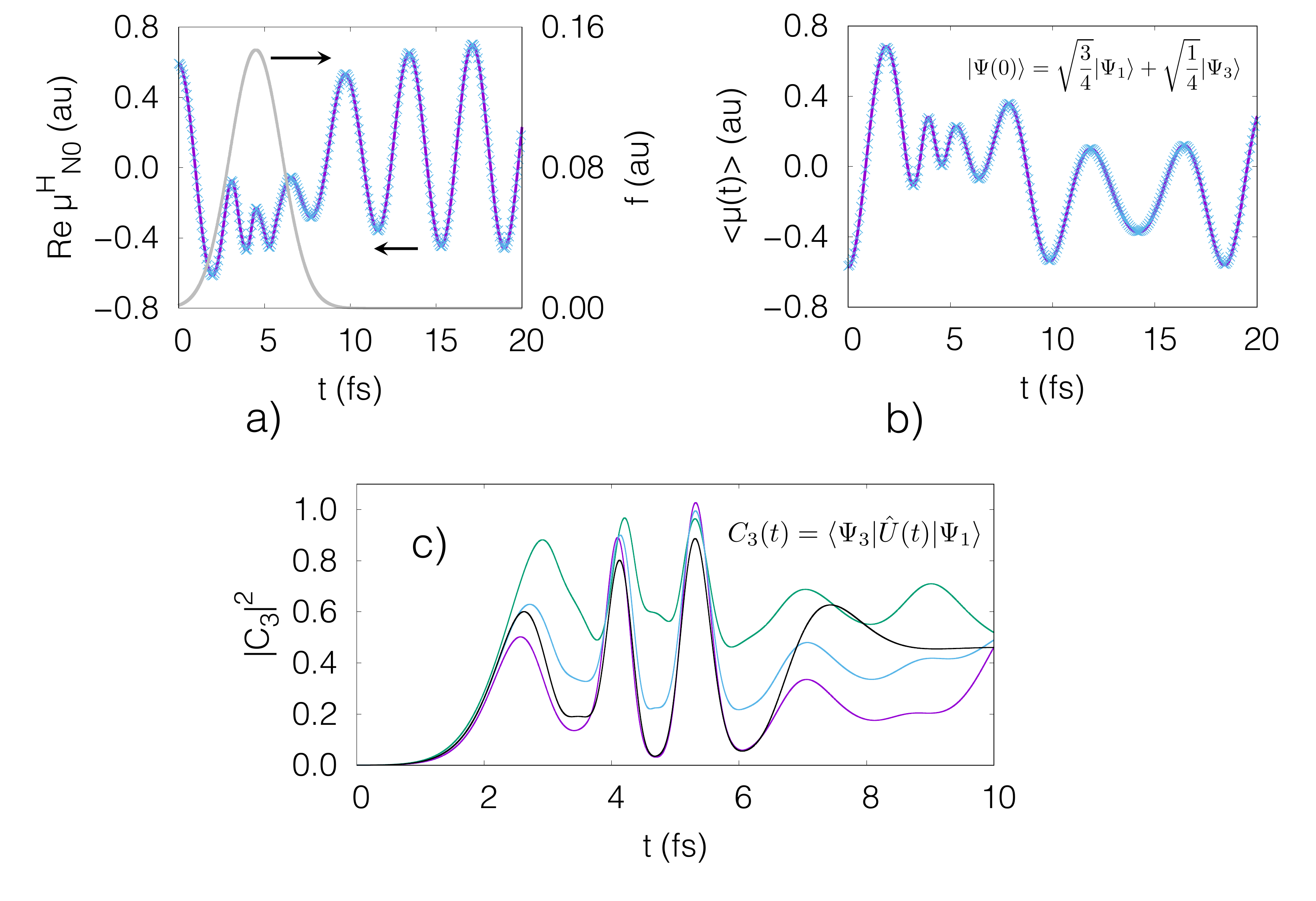}
\caption{Response of the system to a stronger driving pulse in which $f_0\mu_0 = 2$ eV (so
$f_0\approx 7.6\times 10^{10}$ V/m), $\sigma_0 = 1.5$ fs, and $t_0=4.5$ fs. Subfigure a) shows the
shape of the pulse and time-dependency of the real part of the element $\langle
\Psi_N|\hat{\mu}^{\mr{H}}(t)|\Psi_0\rangle$, b) the element $\langle
\Psi(0)|\hat{\mu}^{\mr{H}}|\Psi(0)\rangle$, and c) the evolution of the coefficient $C_3$, and our
CC estimators. Colors and symbols are the same as in Figures \ref{AHH}, \ref{c3square}, and
\ref{APsi}: purple lines refer to SR calculations, blue ``X'' symbols to unitary reference results, and
in c) the purple line is the left-handed estimator, green the right-handed one, blue their average, and
black the exact result.}\label{sstrong}
\end{figure}

The effect of increasing the intensity of the electric field is presented in Fig. \ref{sstrong}
where the unitary propagation results are reproduced for the observable. Despite this, however,
the terms $c_3'$ and $\tilde{d}_{\mr{l},3}^*$ display deviations and an oscillatory behavior at longer times. 
This is caused by the non-Hermitian nature of our time-dependent CC wavefunctions. 
Because the left and right kets are different, there is likely an imbalance in the projections we
extracted from such TD CC kets. However, we believe that with all the tools developed here an
alternative more accurate route to compute eigenstate probabilities may be found, possibly by
analyzing the behavior of the system under different initial conditions.
Non-Hermitian CC theories are the subject of asymmetries that can cause small
deviations from the unitary calculations. The
matrix elements that are inferred from unitary standard quantum mechanics are identified in
non-symmetric non-Hermitian TD CC theory, however, matrix elements from CC do not conjugate as
expected \cite{pedersen1997coupled}, resulting in disparities. In our simulations these are small. There are differences
between the SR CC and the unitary calculations that do not meet the eye, and are below
0.1 \%, but they persist for very fine time grids. For this reason, a potential alternative is to
formulate our theory within unitary coupled-cluster theory, which has quite desirable properties in
terms of the assignment of transition elements. On the other hand, for convenience we employed a
simplified two-electron/two-level which was tuned to feature non-negligible couplings between the
configurations that span the linear space of interest. However, future work could focus on the
application of our initial-state modifications within the context of Lipkin models
\cite{lipkin1965validity,hoodbhoy1978time,harsha2018difference,wahlen2017merging,cervia2021lipkin}, which are
often employed to gain a critical understanding of many-body systems, and may offer in-depth
insights regarding the numerical performance of the proposed methodologies. 

\section{Conclusion}
An extended linear response theory (or second linear response theory) was
formulated to determine properties of excited states through the time-dependent coupled-cluster
formalism, where the generalization to cases beyond that of linear perturbations was considereed. From
the theoretical generalization we derive a set of equations that characterize the time-dependent
evolution of transition elements in the Heisenberg representation, so these could support
propagations that rely on such kind of transition objects or to derive non-linear properties that
rely on linked coupled-cluster expressions. The proposed second response theories can be used to
study quantities such as multipolar matrix elements, magnetic transition amplitudes, and electronic
densities. In the case of second linear response theory, we found it gives results fully consistent
with the well-known coupled-cluster quadratic response theory. On the other hand, because our theory
examines excited states in a step-by-step fashion, it allows us to identify wave-function
time-dependent linear-combination coefficients, so bridging the second linear- and general-response
theory expressions with standard wave function theory. These connections could serve useful in the
computation of excited-state coherent interferences and their response to driving fields in either
the linear or non-linear regime.

\begin{acknowledgments}
M.A.M. acknowledges support by the National Science Foundation through the MonArk Quantum Foundry,
DMR-1906383. The author thanks Prof. Mark A. Ratner (Northwestern University) for motivating early
discussions.
\end{acknowledgments}

\newpage

\renewcommand{\theequation}{S\arabic{equation}}
\begin{center}
{\large \bfseries Supplemental Material}
\end{center}

\section{Standard Response Theory}
In order to derive excited-state quantities, we require regular LR theory to distinguish its
associated properties from excited-state ones. This begins by assuming that the exact full-body
wavefunction of the system is given, which is denoted as $\Psi(t)$.
Thus we consider the following response function:
\ben
R(\omega) = \int_{-\infty}^{+\infty} \ud t~ e^{\ui(\omega\pm \ui\eta) t} \frac{\delta\langle A(t)\rangle }{\delta
f(s)}\Bigg|_{s=0,~f = 0}
\een
\sloppy where $\eta = 0^+$ (this number ensures the integrand decays asymptotically), 
$\langle A(t)\rangle = \langle \Psi(t)|\hat{A}|\Psi(t)\rangle$.
In the ideal case where the exact linear response problem could be solved, one would use
the eigenbasis of the operator $\hat{H}_0$, that is: $\hat{H}_0|\Psi_n\rangle =
E_n|\Psi_n\rangle$, so this spectrum is assumed given as well. 
The standard initial condition for this problem requires that the TD
wavefunction satisfies $|\Psi(t=0)\rangle = |\Psi_0\rangle$, where $\Psi_0$ is the ground
state wavefunction. The TD wavefunction reads $\Psi(t)=\Psi^{(0)}(t)+\Psi^{(1)}(t)$, 
where $\Psi^{(0)}(t) = \exp(-\ui E_0 t)\Psi_0$, and
$\ui \partial_t\Psi^{(1)}(t)=\hat{H}_0\Psi^{(1)}(t)+\hat{v}(t)\Psi^{(0)}(t)$,
$\hat{v}(t)=-f(t)\hat{B}$.

After carrying out the functional derivative with respect to $f(t)$ at $t=0, f=0$, 
and taking the limit $\eta\rightarrow 0^+$, we have that
\ben
R(\omega) = -\sum_n\Bigg[\frac{\langle \Psi_0|\hat{A}|\Psi_n\rangle\langle
\Psi_n|\hat{B}|\Psi_0\rangle}{\omega-\Omega_n}-
\frac{\langle \Psi_0|\hat{B}|\Psi_n\rangle\langle
\Psi_n|\hat{A}|\Psi_0\rangle}{\omega+\Omega_n }\Bigg]
\een
From the poles of the above equation we obtain elements such as $\langle
\Psi_0|\hat{A}_{\mr{N}}|\Psi_n\rangle$ and $\langle \Psi_n|\hat{B}_{\mr{N}}|\Psi_0\rangle$. 

To express $R(\omega)$ using the CC response method, one uses the linearized equations from the main
text, Eqs. (10) and (13). We also express the
(de)excitation amplitudes as 
\ben
\begin{split}
x_{\mu}(t)&=\sum_{I} c_I(t) X_{\mu}^I\\
\lambda_{\mu}(t)&=\sum_{I} d_I(t) \Lambda_{\mu}^I
\end{split}
\een
where $\{c_I(t)\}$ and $\{d_I(t)\}$ are TD complex-valued coefficients.
From the main text Eqs. (10) and (13), through the biorthogonal property we obtain:
\ben
\begin{split}
(\ui\partial_t -\Omega_I)c_I(t)&=\sum_{\mu}
\Lambda^I_{\mu}\langle\hat{\tau}^{\dagger}_{\mu}\bar{v}(t)\rangle_0\\
-(\ui\partial_t +\Omega_I)d_I(t)&=\sum_{\mu}\Big\langle
\hat{L}_0\Big(\bar{v}_{\tau,\mu}(t)+[\bar{H}_{\tau,\mu}^0,\hat{x}(t)]\Big)\Big\rangle_0X^I_{\mu}
\end{split}
\een
For a function $h$ we define the Fourier transform as
$h^{\mr{F}}(\omega)=\int_{-\infty}^{+\infty}\ud t~\exp[\ui(\omega\pm \ui\eta) t]h(t)$, so
$h(t)=(2\pi)^{-1}\int \ud\omega~\exp(-\ui\omega t)h^{\mr{F}}(\omega)$. Furthermore, we can
note that
\ben
\frac{\delta \bar{v}^{\mr{F}}(\omega)}{\delta f(t)}\Bigg|_{t=0} = 
-\bar{B}
\een
After expressing $c_I$ and $d_I$ in Fourier-transformed form, and taking the limit when 
$\eta\rightarrow 0^+$, we obtain that
\ben
\begin{split}
\frac{\delta c_I^{\mr{F}}(\omega)}{\delta f(t)}\Bigg|_{t=0}&=-
\frac{\sum_{\mu}
\Lambda^I_{\mu}\langle\hat{\tau}^{\dagger}_{\mu}\bar{B}\rangle_0}{\omega-\Omega_I}\\
\frac{\delta d_I^{\mr{F}}(\omega)}{\delta f(t)}\Bigg|_{t=0} &=
\frac{\sum_{\mu}
\langle\hat{L}_0\bar{B}_{\tau,\mu}\rangle_0X_{\mu}^I}{\omega+\Omega_I}
+
\sum_{J}
\frac{ 
(\sum_{\mu}\Lambda^J_{\mu}\langle\hat{\tau}_{\mu}^{\dagger}\bar{B}\rangle_0)(\sum_{\mu\nu}X_{\mu}^IF_{\mu\nu}X_{\nu}^J)}
{(\omega+\Omega_I)(\omega-\Omega_J)}
\end{split}
\een
where $F_{\mu\nu}=\langle \hat{L}_0[\bar{H}^0_{\tau,\mu},\hat{\tau}_{\nu}]\rangle_0$, which is a symmetric matrix.

We can express $R(\omega)$ in terms of CC quantities such as:
\ben
R(\omega) = \sum_{\mu,I}\Bigg\{\frac{\delta d^{\mr{F}}_I(\omega)}{\delta
f(t)}\Big|_{t=0,f=0}\Lambda^I_{\mu}\langle\hat{\tau}^{\dagger}_{\mu} \bar{A}\rangle_0
+
\langle \hat{L}_0 \bar{A}_{\tau,\mu}\rangle_0 X^I_{\mu}
\frac{\delta c^{\mr{F}}_I(\omega)}{\delta f(t)}\Big|_{t=0,f=0}
\Bigg\}
\een
This allows us to identify transition elements, Eqs. (16) and (15).

\section{Derivation of Equation 37}

After linearizing $\langle \Upsilon (t)|\hat{A}|\Phi(t)\rangle$, the
TD observable $A(t)=\langle \Upsilon(t)|\hat{A}|\Phi(t)\rangle$ now reads:
\ben\label{ACC}
\begin{split}
A(t)=\langle\hat{L}_0&\bar{A}\rangle_0 + \langle
g_{\mr{L}}\hat{\lambda}^N(t)\bar{A}+g_{\mr{R}}\hat{L}_0[\bar{A},\hat{x}^N(t)]\rangle_0
+\langle \tilde{\lambda}(t)\bar{A}+\hat{L}_0[\bar{A},\tilde{x}(t)]\rangle_0\\
+\langle & g_{\mr{R}}\big(\hat{L}_0\big[[\bar{A},\hat{x}^N(t)],\tilde{x}(t)\big]
+\tilde{\lambda}(t)[\bar{A},\hat{x}^N(t)]\big)+g_{\mr{L}}\hat{\lambda}^N(t)[\bar{A},\tilde{x}(t)]\rangle_0
\end{split}
\een
For a function $h^{\mr{F}}(\omega)$ we introduce the notation:
\ben
h^{\mr{F}}_f(\omega)=\frac{\delta h^{\mr{F}}(\omega)}{\delta f(t)}\Big|_{t=0,f=0}
\een
We are interested in the terms that remain non-zero after multiplication by the factors $(\omega-\Omega_I)$
or $(\omega+\Omega_I)$, and taking the respective limits. Only the third term in angle brackets in
Eq. (\ref{ACC}) contributes to these limits. 
Hence we define the function 
\ben\label{Peq}
P(\omega;g_{\mr{L}}, g_{\mr{R}}) = \Big\langle
\tilde{\lambda}^{\mr{F}}_f(\omega;g_{\mr{L}},g_{\mr{R}})\bar{A}+
\hat{L}_0[\bar{A}, \tilde{x}^{\mr{F}}_f(\omega;g_{\mr{R}})]\Big\rangle_0
\een
Using this we note that:
\ben\label{P1}
\lim_{\omega\rightarrow\Omega_I}\lim_{g_{\mr{L}},g_{\mr{R}}\rightarrow 0}
-(\omega-\Omega_I)\frac{\partial }{\partial g_{\mr{R}}}P(\omega) = 
\langle\Psi_0|\hat{A}|\Psi_I\rangle
\Big(\langle\Psi_I|\hat{B}|\Psi_N\rangle-\delta_{IN}\langle\hat{L}_0\bar{B}\rangle_0\Big)
\een
and
\ben\label{P2}
\lim_{\omega\rightarrow -\Omega_I}\lim_{g_{\mr{L}},g_{\mr{R}}\rightarrow 0} 
(\omega+\Omega_I)\frac{\partial }{\partial g_{\mr{L}}}P(\omega) = \langle\Psi_I|\hat{A}|\Psi_0\rangle
\Big(\langle\Psi_N|\hat{B}|\Psi_I\rangle-\delta_{NI}\langle \hat{L}_0\bar{B}\rangle_0\Big)
\een
To simplify the subsequent expressions we introduce:
$\hat{X}^J=\sum_{\mu}X_{\mu}^J\hat{\tau}_{\mu}$, 
$\hat{\Lambda}^J = \sum_{\mu}\Lambda_{\mu}^J\hat{\tau}^{\dagger}_{\mu}$. Also we define the
commutator: $\Omega_{X,J} = [\hat{\Omega},\hat{X}^J]$, and expand $\tilde{x}_{\mu}(t)$ 
and $\tilde{\lambda}_{\mu}(t)$ as:
\ben
\begin{split}
\tilde{x}_{\mu}(t)&=\sum_{I} \tilde{c}_I(t) X_{\mu}^I\\
\tilde{\lambda}_{\mu}(t)&=\sum_{I} \tilde{d}_I(t) \Lambda_{\mu}^I
\end{split}
\een
where $\{\tilde{c}_I(t)\}$ and $\{\tilde{d}_I(t)\}$ are complex-valued coefficients that depend on
time. These, as the $\{\tilde{x}_{\mu}(t)\}$ and $\{\tilde{\lambda}_{\mu}(t)\}$ coefficients do,
are functions of the driving potential $f(t)$ and the variables $g_{\mr{L}}$ and $g_{\mr{R}}$.
By projecting Eq. (31) onto the basis spanned by $\{\bm{\Lambda}^I\}$ and transforming the result
into frequency space we observe that ($\tilde{c}_{I,f}^{\mr{F}}(\omega)=\delta
\tilde{c}_I^{\mr{F}}(\omega)/\delta f(t)|_{t=0,f=0}$):
\ben
\lim_{g_{\mr{R}}\rightarrow 0}(\omega-\Omega_I)\frac{ \partial}{\partial g_{\mr{R}}}\tilde{c}_{I,f}^{\mr{F}}(\omega) =
-\langle\hat{\Lambda}^{I}\bar{B}_{X,N}\rangle 
+\sum_{J} \langle\hat{\Lambda}^I\big[[\bar{H}_0,\hat{X}^N],\hat{X}^J\big]\rangle_0
 c_{J,f}^{\mr{F}}(\omega-\Omega_N)
\een
Similarly, for the conjugate amplitudes we have that
\ben
\begin{split}
\lim_{g_{\mr{L}},g_{\mr{R}}\rightarrow 0}&-(\omega+\Omega_I)\frac{\partial}{\partial
g_{\mr{R}}}\tilde{d}^{\mr{F}}_{I,f}(\omega;g)=
\sum_{\mu\nu}X^I_{\mu}F_{\mu\nu}X^J_{\nu}\lim_{g_{\mr{R}}\rightarrow
0}\frac{\partial}{\partial g_{\mr{R}}}
\tilde{c}_{J,f}^{\mr{F}}(\omega)
+\Big\langle
-\hat{L}_0[\bar{B}_{X,I},\hat{X}^N]\\
&+\sum_J\Big\{\hat{L}_0\Big[[\bar{H}_{X,I}^0,\hat{X}^N],\hat{X}^J\Big]c_{J,f}^{\mr{F}}(\omega-\Omega_N) 
+d_{J,f}^{\mr{F}}(\omega-\Omega_N)\hat{\Lambda}^J[\bar{H}^0_{X,I},\hat{X}^N]\Big\}\Big\rangle_0
\end{split}
\een
and
\ben
\lim_{g_{\mr{L}},g_{\mr{R}}\rightarrow 0}-(\omega+\Omega_I)\frac{\partial}{\partial
g_{\mr{L}}}\tilde{d}^{\mr{F}}_{I,f}(\omega;g)=
\Big\langle -\hat{\Lambda}^N\bar{B}_{X,I}+\sum_J
\hat{\Lambda}^N[\bar{H}^0_{X,I},\hat{X}^J]c_{J,f}^{\mr{F}}(\omega+\Omega_N)\Big\rangle_0
\een
In the last three equations there are standard linear response quantities, such as
$c_{J,f}^{\mr{F}}(\omega)$ and $d_{J,f}^{\mr{F}}(\omega)$.
Through Eq. (\ref{Peq}), and upon comparison of the last three equations with Eqs. (\ref{P1}) and
(\ref{P2}), we arrive at Eq. (37).

\section{Python code}
\begin{singlespace}
\begin{footnotesize}
\begin{verbatim}
#!/usr/bin/python2.7

from numpy import *
from scipy import linalg

#Definitions
au2ev   = 27.211 #eV
au2angs = 0.529 #angs
au2fs   = 0.0242 #fs

grnd  = 0
sup   = 1
sdown = 2
db    = 3
nlev  = db+1

epsi = 1.0/au2ev
b    = 0.25/au2ev
w    = 0.25/au2ev
sigt = 5./au2fs
t0   = 2.5 * sigt
f0   = 2.0/au2ev
mu0  = 0.5
t_thresh = 1.e-16
time_length = 8.*sigt # was 30
max_t_step = 60000
N_EE = sup #excited state of interest
N_EE_2 = db
C_EE = sqrt(3./4.)
C_EE_2 = sqrt(1./4.)

def f_pulse(t):
  return f0*exp(-0.5*(t-t0)**2./sigt**2.)

#Free Hamiltonian
H0 = zeros((nlev,nlev))
H0[grnd, sup]    = b
H0[grnd, sdown]  = b
H0[grnd, db]     = w
H0[sup, sup]     = epsi/2.
H0[sup, sdown]   = 0.0
H0[sup, db]      = b
H0[sdown, sdown] = epsi/2. #Trick with diagonal
H0[sdown, db]    = b
H0[db, db]       = 2.*epsi/2. #Trick

H0 = H0+H0.transpose()

tau_up   = zeros((nlev, nlev))
tau_down = zeros((nlev, nlev))
tau_db   = zeros((nlev, nlev))

tau_up[sup, grnd]     = 1.
tau_up[db, sdown]     = 1.
tau_down[sdown, grnd] = 1.
tau_down[db, sup]     = 1.
tau_db = dot(tau_up, tau_down)

#Free Hamiltonian diagonalization
Efree, Cf = linalg.eig(H0)
print "\nFull eigenvalues"
print Efree
print ""
Ereal = Efree.real
idx = Ereal.argsort()
Efree = Efree[idx]
Efreal = Ereal[idx]
Cf = Cf[:,idx]

T0 = zeros((nlev)) #first entry is zero
T1 = zeros((nlev))
Lambda_vec = zeros((nlev))
Tmat = zeros((nlev,nlev))
Amat = zeros((nlev,nlev))

def excivec_to_matrix(tvec):
  global tau_up, tau_down, tau_db
  tmat = tvec[sup] * tau_up + tvec[sdown] * tau_down
  tmat += tvec[db] * tau_db
  return tmat

def commutr (A, B):
  return dot(A,B) - dot(B,A)

def Op_transform (OpM, TM):
  out = TM.copy()
  dum = dot(OpM, TM) - dot(TM, OpM)
  out = OpM + dum
  fac = 1.
  for i in xrange(1,4):
    fac = fac*(i+1)
    dum = commutr(dum, TM)
    out += dum/fac
  return out

cc_energy = 0.
H0_T = Amat.copy()
def free_cluster_amps():
  print "Free cluster amplitudes"
  print "Error"
  global T0, T1, Tmat, H0, cc_energy, H0_T
  maxiter = 1000
  for i in xrange(0,maxiter):
    Tmat = excivec_to_matrix(T0)
    H0_T = Op_transform (H0, Tmat)
    T1[sup] = T0[sup] - H0_T[sup, grnd] / epsi
    T1[sdown] = T0[sdown] - H0_T[sdown, grnd] / epsi
    T1[db] = T0[db] - H0_T[db, grnd] / 2. / epsi
    diff_norm = linalg.norm(T1 - T0, 2) / nlev
    print diff_norm
    if diff_norm < t_thresh:
      cc_energy = H0_T[grnd,grnd]
      print "Finished"
      break
    else:
      T0 = T1.copy()

Tau_all = [1., tau_up, tau_down, tau_db]
def A_matrix():
  global Amat, Tau_all
  dum_mat = Amat.copy()
  for mu in xrange(0, nlev):
    for nu in xrange(1, nlev):
      dum_mat = commutr(H0_T, Tau_all[nu])
      Amat[mu,nu] = dum_mat[mu,0]

def find_lambda():
  global Amat, a_submat
  bvec = zeros((nlev - 1))
  hh_mat = Amat[1:nlev, 1:nlev].transpose()
  for mu in xrange(1,nlev):
    bvec[mu-1] = -Amat[0,mu]
  return linalg.solve(hh_mat, bvec)

free_cluster_amps()
print "\nT, CC Energy, Exact eigenvalues"
print T1, cc_energy, Efreal
print ""

A_matrix()
a_submat = Amat[1:nlev, 1:nlev]
print "A_matrix"
print a_submat
ltemp = find_lambda()
print "\nLambda"
print ltemp
print ""
Lambda_vec[1:nlev] = ltemp[0:nlev-1]

Lambda_mat = excivec_to_matrix(Lambda_vec)

Omega_R, X = linalg.eig(a_submat)
Omega_L, L = linalg.eig(a_submat.transpose())
omreal = Omega_R.real
idx = omreal.argsort()
Omega = omreal[idx]
X = X[:,idx]
L = L[:,idx]

for i in xrange(0,nlev-1):
  Cf[:,i+1] *= sign(dot(L[:,i], Cf[1:nlev,i+1]))

for i in xrange(0, nlev-1):
  nfac = sqrt(dot(L[:,i], X[:,i]))
  X[:,i] /= nfac
  L[:,i] /= nfac

print "CC excitation energies"
print cc_energy+Omega
print ""

A_operator = zeros((nlev,nlev), dtype = complex128)
A_operator = mu0 *(tau_up+tau_down)
A_operator += A_operator.transpose()

timevec = linspace(0, time_length, max_t_step)
dt = time_length / max_t_step

psi0_0 = zeros((nlev), dtype = complex128)
psi0_N = zeros((nlev), dtype = complex128)
psi0_0[:] = Cf[:,0]
psi0_N[:] = C_EE * Cf[:,N_EE] + C_EE_2 * Cf[:,N_EE_2]
psi_t_0 = zeros((nlev, max_t_step+1), dtype = complex128)
psi_t_N = zeros((nlev, max_t_step+1), dtype = complex128)
psi_t_0[:,0] = psi0_0
psi_t_N[:,0] = psi0_N
AH_NN = zeros((max_t_step), dtype = complex128)

print "Standard wave function propagation\n"
for i in xrange(0,max_t_step):
  t = timevec[i] + dt/2.
  Vt = - f_pulse(t) * A_operator
  Ht = H0 + Vt
  AH_NN[i] = dot(psi0_N.conjugate(), dot(A_operator, psi0_N))
  psi_t_0[:,i+1] = dot(linalg.expm(-1.j*Ht*dt), psi0_0)
  psi_t_N[:,i+1] = dot(linalg.expm(-1.j*Ht*dt), psi0_N)
  psi0_0 = psi_t_0[:,i+1]
  psi0_N = psi_t_N[:,i+1]
  proj = dot(Cf[:,N_EE+2].conjugate(), psi0_N)
  #print proj.conjugate()*proj

X_t_0 = zeros((nlev-1, max_t_step), dtype = complex128)
X0  = zeros((nlev-1), dtype = complex128)
#X0[:] = 0. #-> initial condition
X_t_0[:,0] = X0[:]
Xmat = zeros((nlev,nlev), dtype = complex128)
xvv = zeros((nlev), dtype = complex128)

print "Standard Response X\n"
t = timevec[0]
Vt = - f_pulse(t) * A_operator
Ht = H0 + Vt
H_transf = Op_transform (Ht, Tmat)
for i in xrange(0,max_t_step-1):
  xvv[sup:nlev] = X0[:]
  Xmat = excivec_to_matrix(xvv)
  dum_mat = Op_transform (H_transf, Xmat)
  X1 = X0[:] -  1.j * dt * dum_mat[sup:nlev,grnd]
  t = timevec[i] + dt
  Vt = - f_pulse(t) * A_operator
  Ht = H0 + Vt
  H_transf = Op_transform (Ht, Tmat)
  xvv[sup:nlev] = X1[:]
  Xmat = excivec_to_matrix(xvv)
  dum_mat2 = Op_transform (H_transf, Xmat) 
  X_t_0[:,i+1] = X0[:] -  0.5j*dt*(dum_mat[sup:nlev,grnd]+dum_mat2[sup:nlev,grnd])
  X0[:] = X_t_0[:,i+1]
  #print "step, amp, ft", i,  X0[2], f_pulse(t)*au2ev

L_t_0 = zeros((nlev-1, max_t_step), dtype = complex128)
L0  = zeros((nlev-1), dtype = complex128)
Ltmp = L0.copy()
L_t_0[:,0] = L0[:]
Lmat = zeros((nlev,nlev), dtype = complex128)

print "Standard Response Lambda\n"
t = timevec[0]
Vt = - f_pulse(t) * A_operator
Ht = H0 + Vt
Xm = X_t_0[:,0].copy()
xvv[sup:nlev] = Xm[:]
Xmat = excivec_to_matrix(xvv)
HX = Op_transform (Ht, Tmat+Xmat)
for i in xrange(0, max_t_step-1):
  xvv[sup:nlev] = L0[:]
  L0mat = excivec_to_matrix(xvv)
  for mu in xrange(1,nlev):
    HX_mu = commutr(HX,Tau_all[mu])
    dum1 = HX_mu + dot(Lambda_mat.transpose(),HX_mu)
    dum2 = dot(L0mat.transpose(), HX_mu)
    Ltmp[mu-1] = L0[mu-1] + 1.j*dt*(dum1[0,0]+dum2[0,0])
  t = timevec[i] + dt
  Vt_2 = - f_pulse(t) * A_operator
  Ht_2 = H0 + Vt
  Xm_2 = X_t_0[:,i+1]
  xvv[sup:nlev] = Xm_2[:]
  Xmat_2 = excivec_to_matrix(xvv)
  HX_2 = Op_transform (Ht, Tmat+Xmat_2)
  xvv[sup:nlev] = Ltmp[:]
  L_tmp_mat = excivec_to_matrix(xvv)
  for mu in xrange(1,nlev):
    HX_mu = commutr(HX,Tau_all[mu])
    HX_mu_2 = commutr(HX_2,Tau_all[mu])
    dum1 = HX_mu + dot(Lambda_mat.transpose(),HX_mu)
    dum1_2 = HX_mu_2 + dot(Lambda_mat.transpose(),HX_mu_2)
    dum2 = dot(L0mat.transpose(), HX_mu)
    dum3 = dot(L_tmp_mat.transpose(), HX_mu_2)
    L_t_0[mu-1,i+1] = L0[mu-1] + 1.j*dt*0.5*(dum1[0,0]+dum1_2[0,0]+dum2[0,0]+dum3[0,0])
  L0[:] = L_t_0[:,i+1]
  Vt[:,:] = Vt_2[:,:]; Ht[:,:] = Ht_2[:,:]
  Xm[:] = Xm_2[:]; Xmat[:,:] = Xmat_2[:,:]
  HX[:,:] = HX_2[:,:]

L_t = zeros((nlev-1, max_t_step), dtype = complex128) #Lambda_l
L0  = zeros((nlev-1), dtype = complex128)
L0[:] = C_EE * L[:, N_EE-1] + C_EE_2 * L[:, N_EE_2-1]
Xr_t = zeros((nlev-1, max_t_step), dtype = complex128)
X0  = zeros((nlev-1), dtype = complex128)
X0[:] = C_EE * X[:,N_EE-1] + C_EE_2 * X[:,N_EE_2-1]
Xtmp = L0.copy()
Xr_t[:,0] = X0[:]

AH_NN_R = zeros((max_t_step), dtype = complex128)
delta_phi_l = zeros((max_t_step), dtype = complex128)

AH_file = open("AH_NN_file.dat", "w+")
AH_file_2 = open("AH_NN_file_2.dat", "w+")

#F matrix
Lr0  = zeros((nlev-1), dtype = complex128)
Lr_t = zeros((nlev-1, max_t_step), dtype = complex128)
FM = zeros((nlev-1, nlev-1))
for mu in xrange(1,nlev):
  for nu in xrange(1,nlev):
    H0mu = commutr(Op_transform(H0, Tmat), Tau_all[mu])
    dum0 = commutr(H0mu, Tau_all[nu])
    dum0 = dum0 + dot(Lambda_mat.transpose(), dum0)
    FM[mu-1, nu-1] = dum0[0,0]


H0_transf = Op_transform(H0, Tmat)

coeffs = [C_EE, C_EE_2]
for J in xrange(0,nlev-1):
  i = 0; j = 0
  for M in [N_EE-1, N_EE_2-1]:
    j = 0
    for N in [N_EE-1, N_EE_2-1]:
      print i,j
      print M,N
      xvv[sup:nlev] = L[:,M]
      LM = excivec_to_matrix(xvv)
      xvv[sup:nlev] = X[:,N]
      XN = excivec_to_matrix(xvv)
      xvv[sup:nlev] = X[:,J]
      XJ = excivec_to_matrix(xvv)
      dum = dot(LM.transpose(), commutr(commutr(H0_transf, XN), XJ)) / (Omega[M]-Omega[J]-Omega[N])
      Lr0[:] += coeffs[i] * coeffs[j] * dum[0,0] * L[:,J]
      j += 1
    i += 1

#Lr0[:] = 0.
Lr_t[:,0] = Lr0[:]

t = timevec[0]
Vt = - f_pulse(t) * A_operator
Ht = H0+Vt
V_transf = Op_transform(Vt, Tmat)
H_transf = Op_transform(Ht, Tmat)
Xm[:] = X_t_0[:,0]
print Xm
xvv[sup:nlev] = Xm[:]
Xmat = excivec_to_matrix(xvv)
AX = Op_transform(A_operator, Xmat+Tmat)
#VX = Op_transform(V_transf, Xmat)
HX = Op_transform(H_transf, Xmat)

dum = dot(Lambda_mat.transpose(), Op_transform(A_operator, Tmat))
dum +=  Op_transform(A_operator, Tmat)
ANN0 = dum[0,0]

print "Second Response\n"
for i in xrange(0,max_t_step - 1):
#<N|AH|N>
  xvv[sup:nlev] = L0[:]
  LL = excivec_to_matrix(xvv)
  dum = dot(LL.transpose(), AX)
  xvv[sup:nlev] = X0[:]
  Xr_mat = excivec_to_matrix(xvv)
  xvv[sup:nlev] = Lr0[:]
  Lr0mat = excivec_to_matrix(xvv)
  dum1 = dot(LL.transpose(), commutr(AX, Xr_mat))
  dum2 = dot(Lr0mat.transpose(), AX)
  AH_NN_R[i] = dum1[0,0] + dum2[0,0]
  xvv[sup:nlev] = L_t_0[:,i]
  L0_mat = excivec_to_matrix(xvv)  #Lambda(t) from gs propagation
  dum = AX + dot(Lambda_mat.transpose(), AX)
  dum += dot(L0_mat.transpose(), AX)
  AH00 = dum[0,0]
  AH_NN_R[i] += AH00

#midpoint algo
  t = timevec[i] + dt
  Vt_2 = - f_pulse(t) * A_operator
  Ht_2 = H0+Vt_2
  V_transf_2 = Op_transform(Vt_2, Tmat)
  H_transf_2 = Op_transform(Ht_2, Tmat)

#Lambda_l
  Xm[:] = X_t_0[:,i+1]
  xvv[sup:nlev] = Xm[:]
  Xmat_2 = excivec_to_matrix(xvv)
  AX_2 = Op_transform(A_operator, Xmat_2+Tmat)
  HX_2 = Op_transform(H_transf_2, Xmat_2)
  xvv[sup:nlev] = L0[:]
  L_tmp_mat = excivec_to_matrix(xvv)
  for mu in xrange(1,nlev):
    dum1 = commutr(HX,Tau_all[mu])
    dum1 = dot(L_tmp_mat.transpose(), dum1)
    Ltmp[mu-1] = L0[mu-1] + 1.j*dt*dum1[0,0]
  xvv[sup:nlev] = Ltmp[:]
  L_tmp_mat_2 = excivec_to_matrix(xvv)
  for mu in xrange(1,nlev):
    dum1 = commutr(HX,Tau_all[mu])
    dum1 = dot(L_tmp_mat.transpose(), dum1)
    dum2 = commutr(HX_2,Tau_all[mu])
    dum2 = dot(L_tmp_mat_2.transpose(), dum2)
    L_t[mu-1,i+1] = L0[mu-1] + 1.j*dt*0.5*(dum1[0,0]+dum2[0,0])
  L0[:] = L_t[:,i+1]

#Xr
  xvv[sup:nlev] = X0[:]
  X0_mat = excivec_to_matrix(xvv)
  dum1 = commutr(HX, X0_mat)
  X1 = X0[:] - 1.j*dt*dum1[sup:nlev,0]
  xvv[sup:nlev] = X1[:]
  X1_mat = excivec_to_matrix(xvv)
  dum2 = commutr(HX_2, X1_mat)
  Xr_t[:,i+1] = X0[:] - 1.j*dt*0.5*(dum1[sup:nlev,0]+dum2[sup:nlev,0])
  X0[:] = Xr_t[:,i+1]

#lambda_lr
  xvv[sup:nlev] = Xr_t[:,i]
  Xr_mat = excivec_to_matrix(xvv)
  xvv[sup:nlev] = Xr_t[:,i+1]
  Xr_mat_2 = excivec_to_matrix(xvv)
  xvv[sup:nlev] = Lr0[:]
  Lr0mat = excivec_to_matrix(xvv)
  xvv[sup:nlev] = L0[:]
  LL_2 = excivec_to_matrix(xvv)
  for mu in xrange(1,nlev):
    HXmu = commutr(HX, Tau_all[mu])
    dum1 = dot(LL.transpose(), commutr(HXmu, Xr_mat))
    dum2 = dot(Lr0mat.transpose(), HXmu)
    Ltmp[mu-1] = Lr0[mu-1] + 1.j*dt*(dum1[0,0]+dum2[0,0])
  xvv[sup:nlev] = Ltmp[:]
  Lr0mat_2 = excivec_to_matrix(xvv)
  for mu in xrange(1,nlev):
    HXmu = commutr(HX, Tau_all[mu])
    HXmu_2 = commutr(HX_2, Tau_all[mu])
    dum1 = dot(LL.transpose(), commutr(HXmu, Xr_mat))
    dum1 += dot(LL_2.transpose(), commutr(HXmu_2, Xr_mat_2))
    dum2 = dot(Lr0mat.transpose(), HXmu)
    dum2 += dot(Lr0mat_2.transpose(), HXmu_2)
    Lr_t[mu-1,i+1] = Lr0[mu-1] + 1.j*dt*0.5*(dum1[0,0]+dum2[0,0])
  Lr0[:] = Lr_t[:,i+1]

  time_fs = timevec[i]*0.0242
  print >> AH_file, time_fs, AH_NN_R[i].real, AH_NN[i].real
  print >> AH_file_2, time_fs, AH_NN_R[i].imag, AH_NN[i].imag
  print AH_NN_R[i], AH_NN[i]

  Vt[:,:] = Vt_2[:,:]; Ht[:,:] = Ht_2[:,:]
  V_transf[:,:] = V_transf_2[:,:]; H_transf[:,:] = H_transf_2[:,:]
  Xm[:] = Xm_2[:]; Xmat[:,:] = Xmat_2[:,:]
  AX[:,:] = AX_2[:,:]; HX[:,:] = HX_2[:,:]


AH_file.close(); AH_file_2.close()

print "Done"

\end{verbatim}
\end{footnotesize}
\end{singlespace}

\end{document}